\documentclass{aa}
\usepackage{txfonts}
\usepackage{graphicx}
\usepackage{caption}
\usepackage{subcaption}
\usepackage{color}
\usepackage{tablefootnote}

\usepackage{tikz}
\usetikzlibrary{arrows,shapes,backgrounds}

\begin{document} 
\titlerunning{}
\authorrunning{Mountrichas et al.}
\titlerunning{Star formation of X-ray AGN in COSMOS: The role of AGN activity and galaxy stellar mass}

\title{Star formation of X-ray AGN in COSMOS: The role of AGN activity and galaxy stellar mass}

\author{G. Mountrichas\inst{1}, V. A. Masoura\inst{1,2},  E.M. Xilouris\inst{2}, I. Georgantopoulos\inst{2},   V. Buat\inst{3,4}, E.-D. Paspaliaris\inst{2,5}}
          
     \institute {Instituto de Fisica de Cantabria (CSIC-Universidad de Cantabria), Avenida de los Castros, 39005 Santander, Spain
              \email{gmountrichas@gmail.com}
           \and
             National Observatory of Athens, Institute for Astronomy, Astrophysics, Space Applications and Remote Sensing, Ioannou Metaxa
and Vasileos Pavlou GR-15236, Athens, Greece
	 \and
             Aix Marseille Univ, CNRS, CNES, LAM Marseille, France. 
                \email{ veronique.buat@lam.fr}  
              \and
                 Institut Universitaire de France (IUF)
                 \and
                 Department of Astrophysics, Astronomy \& Mechanics, School of Physics, Aristotle University of Thessaloniki, 54124 Thessaloniki, Greece
        }

\abstract {We use $\sim 1,000$ X-ray sources in the {\it{COSMOS-Legacy}} survey and study the position of the AGN relative to the star forming main sequence (MS). We also construct a galaxy (non-AGN) reference sample that includes $\sim 90,000$ sources. We apply the same photometric selection criteria on both datasets and construct their spectral energy distributions (SEDs) using optical to far-infrared photometry compiled by the HELP project. We perform SED fitting, using the X-CIGALE algorithm and the same parametric grid for both datasets, to measure the star formation rate (SFR) and stellar mass of the sources. The mass completeness of the data is calculated at different redshift intervals and is applied on both samples. We define our own main sequence, based on the distributions of the specific SFR at different redshift ranges and exclude quiescent galaxies from our analysis. These allow us to compare the SFR of the two populations in a uniform manner, minimizing systematics and selection effects. Our results show that at low to moderate X-ray luminosities, AGN tend to have lower or, at most, equal star formation rates compared to non-AGN systems with similar stellar mass and redshift. At higher ($\rm L_{X,2-10keV} > 2-3\times 10^{44}\,erg\,s^{-1}$), we observe an increase of the SFR of AGN, for systems that have $\rm 10.5 < log\,[M_*(M_\odot)] < 11.5$.} 

   
\maketitle

\section{Introduction}

Active galactic nuclei (AGN) are powered by accretion onto supermassive black holes (SMBHs) that are located in the centre of galaxies. However, the exact mechanism(s) that triggers their activity is still elusive. When the SMBH becomes active it releases enormous amounts of energy, known as AGN feedback. Although, orders of magnitude in scale separate galaxies from their SMBHs, it is widely accepted that AGN feedback plays an important role in the life and evolution of the entire host galaxy \citep[e.g.,][]{Hickox2018}. AGN feedback, in the form of jets, radiation or winds is included in most simulations, to explain many galaxy properties, such as to maintain the hot intracluster medium \citep[ICM; e.g.,][]{Dunn2006} explain the shape of the galaxy stellar mass function \citep[e.g.][]{Bower2012} and the galaxy morphology \citep[e.g.][]{Dubois2016}. Different mechanisms have been suggested to drive this energetic outflows from the SMBH to the galaxy spheroid \citep[for a review see][]{Morganti2017}.

One aspect of this AGN-galaxy co-evolution is whether there is a link between the activity of the black hole and the star formation (SF) of the host galaxy. X-rays detect the activity of the central SMBH, and therefore the X-ray luminosity, L$_X$, is used as a proxy of the AGN power. Although many works have studied the SF of the galaxy as a function of L$_X$, this is still a challenging task. Initial studies were hampered by cosmic variance and low number statistics \citep[e.g.,][]{Lutz2010, Page2012}. In more recent years, larger samples were used \citep[e.g.][]{Lanzuisi2017}, but it also became apparent that the SFR-L$_X$ relation alone does not provide many useful information. More insights can be gained by comparing the SFR of AGN host galaxies with the SFR of non-AGN systems with similar properties (M$_*$, redshift). An additional complication comes from the fact that X-ray AGN (as opposed e.g. to optical QSOs) span approximately four orders of magnitude in luminosity. Thus, it may be possible that the SFR-L$_X$ relation changes at different L$_X$ regimes.  

One popular method to compare the SFR of AGN with that of non-AGN systems, is to adopt for the calculation of the latter, analytical expressions from the literature, that describe the SFR-M$_*$ correlation, known as main sequence \citep[MS; e.g.,][]{Noeske2007, Elbaz2007, Whitaker2012, Speagle2014}. The estimated parameter is the SFR$_{norm}$, defined as the ratio of the SFR of AGN to the SFR of MS galaxies. Results from studies that followed this approach \citep[e.g.,][]{Mullaney2015, Masoura2018, Bernhard2019, Masoura2021, Torbaniuk2021}, though, may suffer from systematics. These systematic could be introduced by a number of factors. For example,  different methods are applied for the estimation of the host galaxy properties (SFR, M$_*$) of AGN and that of star forming (non-AGN) systems, the definition of MS is not strict and different selection criteria have been applied on the AGN and non-AGN galaxy samples. 

A similar, but improved approach, is to compare the SFR of AGN with that from a control galaxy, i.e., a non-AGN sample that has been selected by applying the same criteria (e.g. photometric coverage) as the AGN sample and for which the galaxy properties have been calculated following the same method (e.g. SED fitting). \cite{Shimizu2015, Shimizu2017} used ultra-hard X-ray selected AGN from the {\it{Swift}} Burst Alert Telescope (BAT) at $\rm z<0.1$, and studied the location of AGN in the SFR-M$_*$ plane. For that purpose, they defined their own MS, using galaxies for which they have applied comparable methods to estimate their SFR and M$_*$, as for the AGN population. Their analysis showed that a large fraction of AGN lie below the MS. A mild dependence of SFR with L$_X$ is detected with a large scatter. \cite{Florez2020} used X-ray AGN in the Stripe 82 field and compared their SFRs with non-X-ray systems. Their results showed that AGN tend to have three to ten times higher SFR compared to the non-X-ray sources, at the same stellar mass and redshift. More recently, \cite{Mountrichas2021c}, used X-ray AGN in the Bo$\rm \ddot{o}$tes field and found that SFR$_{norm}$ does not evolve with redshift. Their results also suggest, that in less massive galaxies ($\rm log\,[M_*(M_\odot)] \sim 11$), AGN hosts have enhanced SFR by $\sim 50\%$ compared to non-AGN systems. A flat relation is observed for the most massive galaxies.

The COSMOS field \citep{Scoville2007} offers a unique combination of deep and multiwavelength data, from radio to X-rays, in an area of about 2\,deg$^2$. This plethora of data combined with the available X-ray observations in this field, make it ideal to accurately measure and compare the galaxy properties of X-ray AGN and non-AGN systems. \cite{Santini2012} used X-ray selected AGN in different fields, including the XMM-COSMOS \citep{Cappelluti2009}, and compared their average SFR with that of a mass-matched control sample of non-AGN galaxies, in the $\rm 0.5<z<2.5$ redshift range. Their analysis showed that AGN present an enhanced FIR emission with respect to inactive galaxies of similar mass. However, the locus of AGN hosts is broadly consistent with the MS of only star forming galaxies (see their Figure 5). \cite{Rosario2013} used optically selected and X-ray detected QSOs in the COSMOS field and compared their SFR with normal massive star forming galaxies, using the equation of \cite{Whitaker2012}. They found that the mean SFR of QSOs are consistent with those of star forming galaxies. \cite{Lanzuisi2017}, used $\sim 700$ X-ray AGN in the COSMOS field and found a significant correlation between the X-ray and star formation luminosities. However, when they binned and averaged the two quantities, they found that the observed trends depend on which parameter is binned \citep[e.g., see also][]{Masoura2021}. A plausible interpretation of this behaviour is that SFR is a slowly changing galaxy property as opposed to the rapidly changing L$_X$ \citep{Hickox2014, Volonteri2015}. \cite{Bernhard2019} used X-ray sources in the Chandra COSMOS area \citep{Marchesi2016}, in the $\rm 0.8<z<1.2$ redshift range. They estimated the SFR$_{norm}$ parameter, using equation 9 of \cite{Schreiber2015} to calculate the SFR of star-forming MS galaxies. They found that AGN with L$_X>2 \times 10^{43}\rm \,erg s^{-1}$ have a narrower SFR$_{norm}$ distribution that is shifted to higher values compared to their lower L$_X$ counterparts.

In this work, we utilize the abundance of data availability in the {\it{Chandra}} COSMOS field and compare the SFR of X-ray selected AGN with those from a galaxy control sample. We follow the method presented in \cite{Mountrichas2021c} to estimate the SFR$_{norm}$ parameter. Our main goal is to extend the luminosity baseline presented in the Mountrichas et al. work to an order of magnitude lower X-ray luminosities, following a uniform methodology and applying similar photometric criteria to select AGN and non-AGN systems with them. Moreover, we will examine whether we confirm their findings of enhanced SFR for AGN at   L$_X>10^{44}\rm \,erg s^{-1}$ for systems with $\rm log\,[M_*(M_\odot)] \sim 11$. In Sect. \ref{sec_data} we describe the AGN and non-AGN samples and the available photometry. Sect. \ref{sec_analysis} presents the SED fitting analysis we perform to calculate the galaxy properties of our sources, the mass completeness of the data and our definition of MS. The SFR$_{norm}-$L$_X$ relation is studied in Sect. \ref{sec_lx_sfr}. In Sect. \ref{sec_summary} we summarise the results of our analysis.

Throughout this work, we assume a flat $\Lambda$CDM cosmology with $H_ 0=69.3$\,Km\,s$^{-1}$\,Mpc$^{-1}$ and $\Omega _ M=0.286$.

\begin{figure}
\centering
\begin{subfigure}{.5\textwidth}
  \centering
  \includegraphics[width=1.\linewidth, height=7cm]{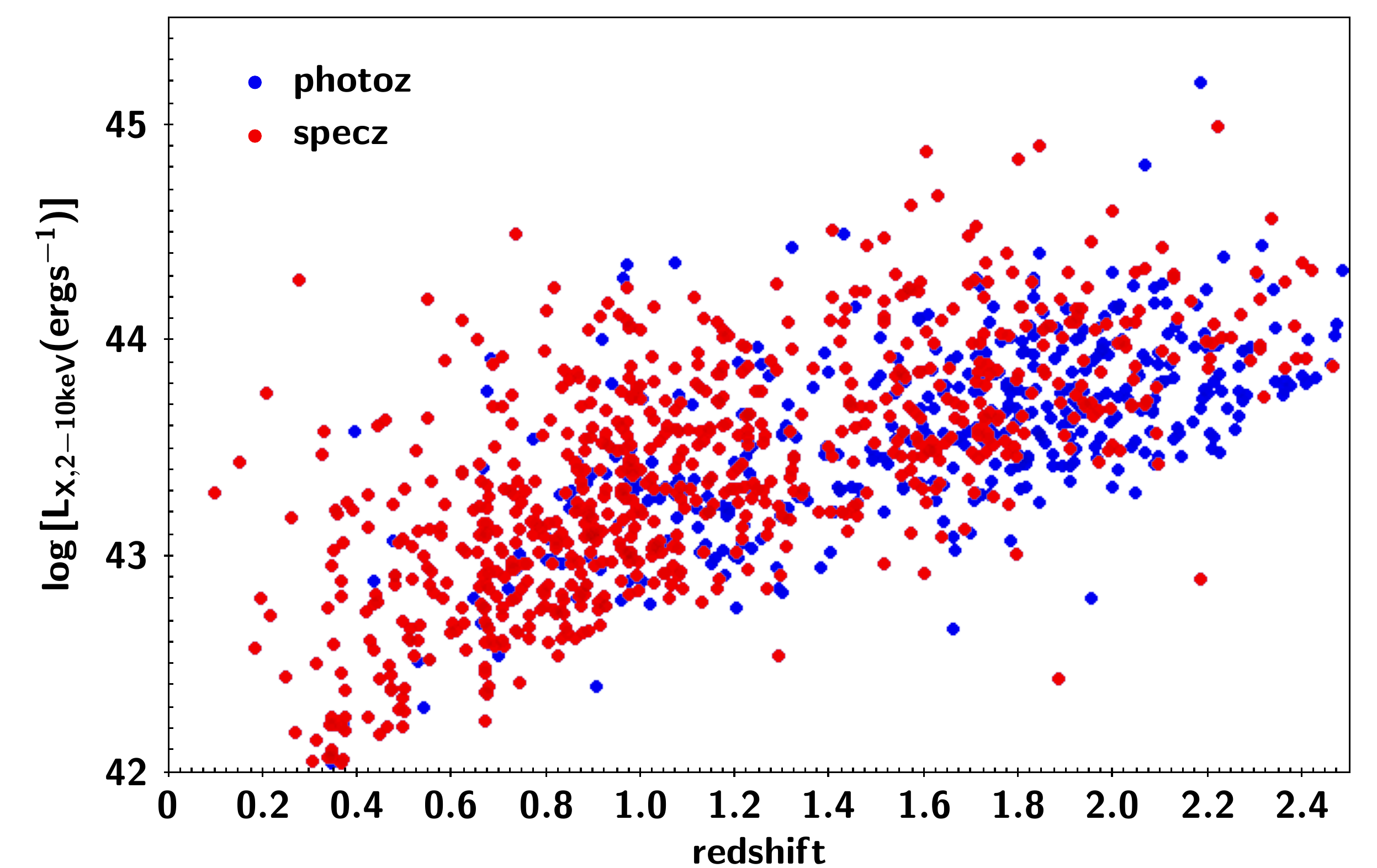}
  \label{}
\end{subfigure}
\begin{subfigure}{.5\textwidth}
  \centering
  \includegraphics[width=1.\linewidth, height=7cm]{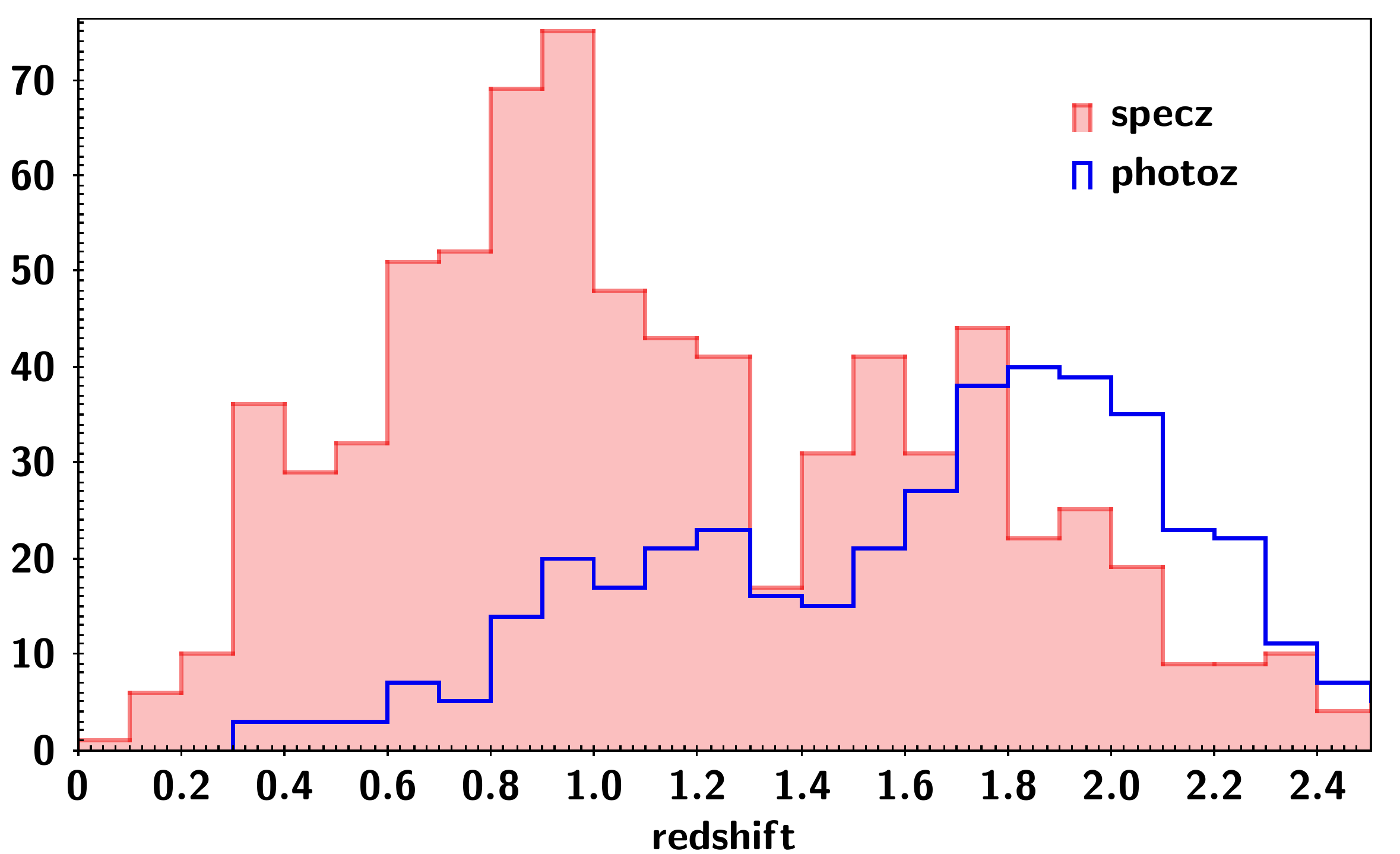}
  \label{}
  \end{subfigure}
  \caption{Top panel: X-ray luminosity as a function of redshift for the 1,236 X-ray AGN. Red circles present sources with spectroscopic redshifts available and blue circles those with photometric redshifts. Bottom panel: Redshift distributions of photoz and specz sources. The majority of X-ray sources have specz at $\rm z<1$, whereas at $\rm z>1.5$ most of the sources have photometric redshifts.}
  \label{fig_lx_redz}
\end{figure}

\section{Data}
\label{sec_data}

\subsection{X-ray sample}
The {\it{COSMOS-Legacy}} survey \citep{Civano2016} is a 4.6\,Ms {\it{Chandra}} program that covers 2.2\,deg$^2$ of the COSMOS field \citep{Scoville2007}. The central area has been observed with an exposure time of $\approx 160$\,ks while the remaining area has exposure time of $\approx 80$\,ks. The limiting depths are $2.2 \times 10^{-16}$, $1.5 \times 10^{-15}$ and $8.9 \times 10^{-16}\,\rm erg\,cm^{-2}\,s^{-1}$ in the soft (0.5-2\,keV), hard (2-10\,keV) and full (0.5-10\,keV) bands, respectively. The X-ray catalogue includes 4016 sources. \cite{Marchesi2016} matched the X-ray sources with optical/infrared counterparts, using the likelihood ratio technique \citep{Sutherland_and_Saunders1992}. 97\% of the sources have an optical/IR counterpart and a photometric redshift (photoz) and $\approx 54\%$ of the sources have spectroscopic redshift (specz). Photometric redshifts have been produced following the procedure described in \cite{Salvato2011}. Different libraries of templates have been used, depending on the X-ray flux of the source and the morphological and photometric properties of the associated counterpart. The best fit has been derived using the LePhare code \citep{Arnouts1999, Ilbert2006}. The accuracy of photometric redshifts is found at $\sigma_{\Delta z/(1+z_{spec})}=0.03$. The fraction of outliers ($\Delta z/(1+z_{zspec})>0.15$) is $\approx 8\%$. Hardness ratios ($\rm HR=\frac{H-S}{H+S}$, where H and S are the net counts of the sources in the hard and in the soft band, respectively) have been estimated for all X-ray sources, using the Bayesian Estimation of Hardness Ratios method \citep[BEHR;][]{Park2006}. The intrinsic column density, N$\rm _H$ for each source has then been calculated, using its redshift and assuming an X-ray spectral power-law with slope $\Gamma=1.8$. This information is available in the catalogue presented in \cite{Marchesi2016}.

\begin{table*}
\caption{The models and the values for their free parameters used by X-CIGALE for the SED fitting.} 
\centering
\setlength{\tabcolsep}{1.mm}
\begin{tabular}{cc}
       \hline
Parameter &  Model/values \\
	\hline
\multicolumn{2}{c}{Star formation history: delayed model and recent burst} \\
Age of the main population & 1500, 2000, 3000, 4000, 5000, 7000, 10000, 12000 Myr \\
e-folding time & 200, 500, 700, 1000, 2000, 3000, 4000, 5000 Myr \\ 
Age of the burst & 50 Myr \\
Burst stellar mass fraction & 0.0, 0.005, 0.01, 0.015, 0.02, 0.05, 0.10, 0.15, 0.18, 0.20 \\
\hline
\multicolumn{2}{c}{Simple Stellar population: Bruzual \& Charlot (2003)} \\
Initial Mass Function & Chabrier (2003)\\
Metallicity & 0.02 (Solar) \\
\hline
\multicolumn{2}{c}{Galactic dust extinction} \\
Dust attenuation law & Charlot \& Fall (2000) law   \\
V-band attenuation $A_V$ & 0.2, 0.3, 0.4, 0.5, 0.6, 0.7, 0.8, 0.9, 1, 1.5, 2, 2.5, 3, 3.5, 4 \\ 
\hline
\multicolumn{2}{c}{Galactic dust emission: Dale et al. (2014)} \\
$\alpha$ slope in $dM_{dust}\propto U^{-\alpha}dU$ & 2.0 \\
\hline
\multicolumn{2}{c}{AGN module: SKIRTOR)} \\
Torus optical depth at 9.7 microns $\tau _{9.7}$ & 3.0, 7.0 \\
Torus density radial parameter p ($\rho \propto r^{-p}e^{-q|cos(\theta)|}$) & 1.0 \\
Torus density angular parameter q ($\rho \propto r^{-p}e^{-q|cos(\theta)|}$) & 1.0 \\
Angle between the equatorial plan and edge of the torus & $40^{\circ}$ \\
Ratio of the maximum to minimum radii of the torus & 20 \\
Viewing angle  & $30^{\circ}\,\,\rm{(type\,\,1)},70^{\circ}\,\,\rm{(type\,\,2)}$ \\
AGN fraction & 0.0, 0.1, 0.2, 0.3, 0.4, 0.5, 0.6, 0.7, 0.8, 0.9, 0.99 \\
Extinction law of polar dust & SMC \\
$E(B-V)$ of polar dust & 0.0, 0.2, 0.4 \\
Temperature of polar dust (K) & 100 \\
Emissivity of polar dust & 1.6 \\
\hline
\multicolumn{2}{c}{X-ray module} \\
AGN photon index $\Gamma$ & 1.4 \\
Maximum deviation from the $\alpha _{ox}-L_{2500 \AA}$ relation & 0.2 \\
LMXB photon index & 1.56 \\
HMXB photon index & 2.0 \\
\hline
Total number of models (X-ray/reference galaxy catalogue) & 313,632,000/60,984,000 \\
\hline
\label{table_cigale}
\end{tabular}
\tablefoot{For the definition of the various parameter see section \ref{sec_cigale}.}
\end{table*}

In our analysis, we use only sources within both the COSMOS and UltraVISTA \citep{McCracken2012}  regions. UltraVISTA covers 1.38\,deg$^2$ of the COSMOS field \citep[after removing the masked objects, see Fig. 1 in][]{Laigle2016} and has deep near infrared observations ($J, H, K_s$ photometric bands) that will allow us to derive more accurate host galaxy properties through SED fitting (see below). There are 1,718 X-ray sources that lie within the UltraVISTA area of COSMOS. From them, 1,627 satisfy the photometric criteria (see next paragraph) and 1,236 also meet our reliability requirements (see section \ref{sec_bad_fits}). The X-ray luminosity as a function of redshift for the 1,236 X-ray sources is presented in the top panel of Fig. \ref{fig_lx_redz}. AGN with specz (809 sources) are shown in red, whereas AGN with photoz (427 sources) are shown in blue. The majority of X-ray sources at $\rm z<1$ have specz, whereas at $\rm z>1.5$ most of the sources have photoz (bottom panel of Fig. \ref{fig_lx_redz}).

The X-ray catalogue is cross-matched with the COSMOS photometric catalogue produced by the HELP collaboration \citep{Shirley2019, Shirley2021}. HELP includes data from 23 of the premier extragalactic survey fields, imaged by the {\it{Herschel}} Space Observatory which form the {\it{Herschel}} Extragalactic Legacy Project (HELP). The catalogue provides homogeneous and calibrated multiwavelength data. The cross-match with the HELP catalogue is done using 1'' radius and the optical coordinates of the counterpart of each X-ray source. In our analysis, we need reliable estimates of the galaxy properties via SED fitting. Therefore, we require all our X-ray AGN to have been detected in the following photometric bands $u, g, r, i, z, J, H, K_s$, IRAC1, IRAC2 and MIPS/24. IRAC1, IRAC2 and MIPS/24 are the [3.6]\,$\mu$m, [4.5]\,$\mu$m and 24 microns, photometric bands of Spitzer.

\subsection{Galaxy reference catalogue}

In our analysis, we compare the SFR of X-ray AGN with that of normal, non-AGN, galaxies. To perform a fully consistent comparison between the SFR of AGN and non-AGN systems, we apply the same SED fitting analysis in both datasets and we require the same availability of photometric bands. The reference catalogue is provided by the HELP collaboration{\footnote{HELP dataset includes sources from the catalogue presented in \cite{Laigle2016}. A new catalogue \citep[COSMOS2020;][]{Weaver2021} was recently released. However, this paper was already in an advanced stage when the new catalogue was made available.}} . There are about $\sim 2.5$ million galaxies in the COSMOS field. $\sim 500,000$ are in the 1.38\,deg$^2$ of UltraVISTA \citep[see also][]{Laigle2016}. There are $\sim 230,000$ galaxies, after we exclude X-ray sources, that meet the photometric requirements we have set on the X-ray sample.


\begin{figure*}
\centering
\begin{subfigure}{.45\textwidth}
  \centering
  \includegraphics[width=1.\linewidth, height=7cm]{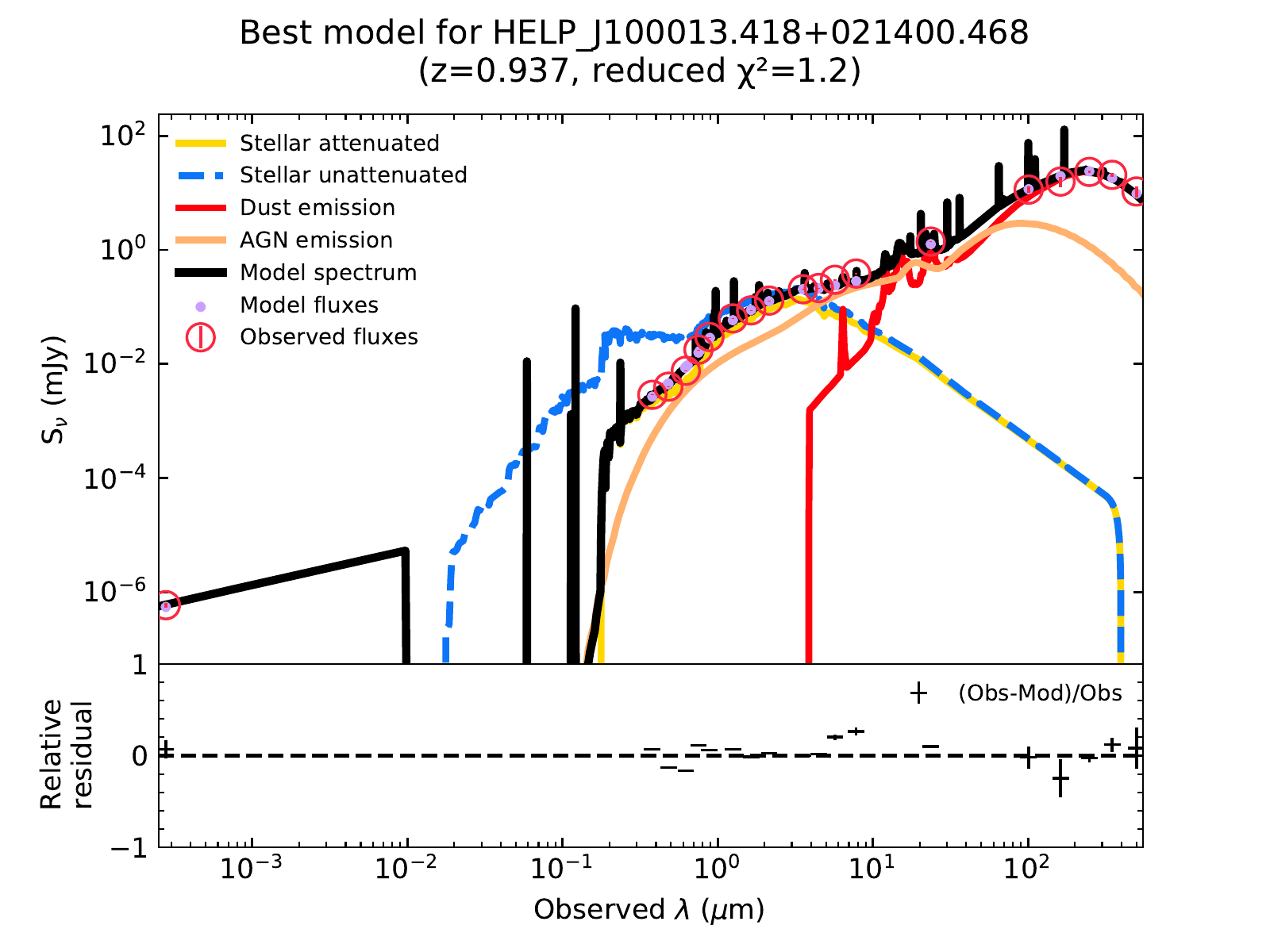}
  \label{}
\end{subfigure}
\begin{subfigure}{.45\textwidth}
  \centering
  \includegraphics[width=1.\linewidth, height=7cm]{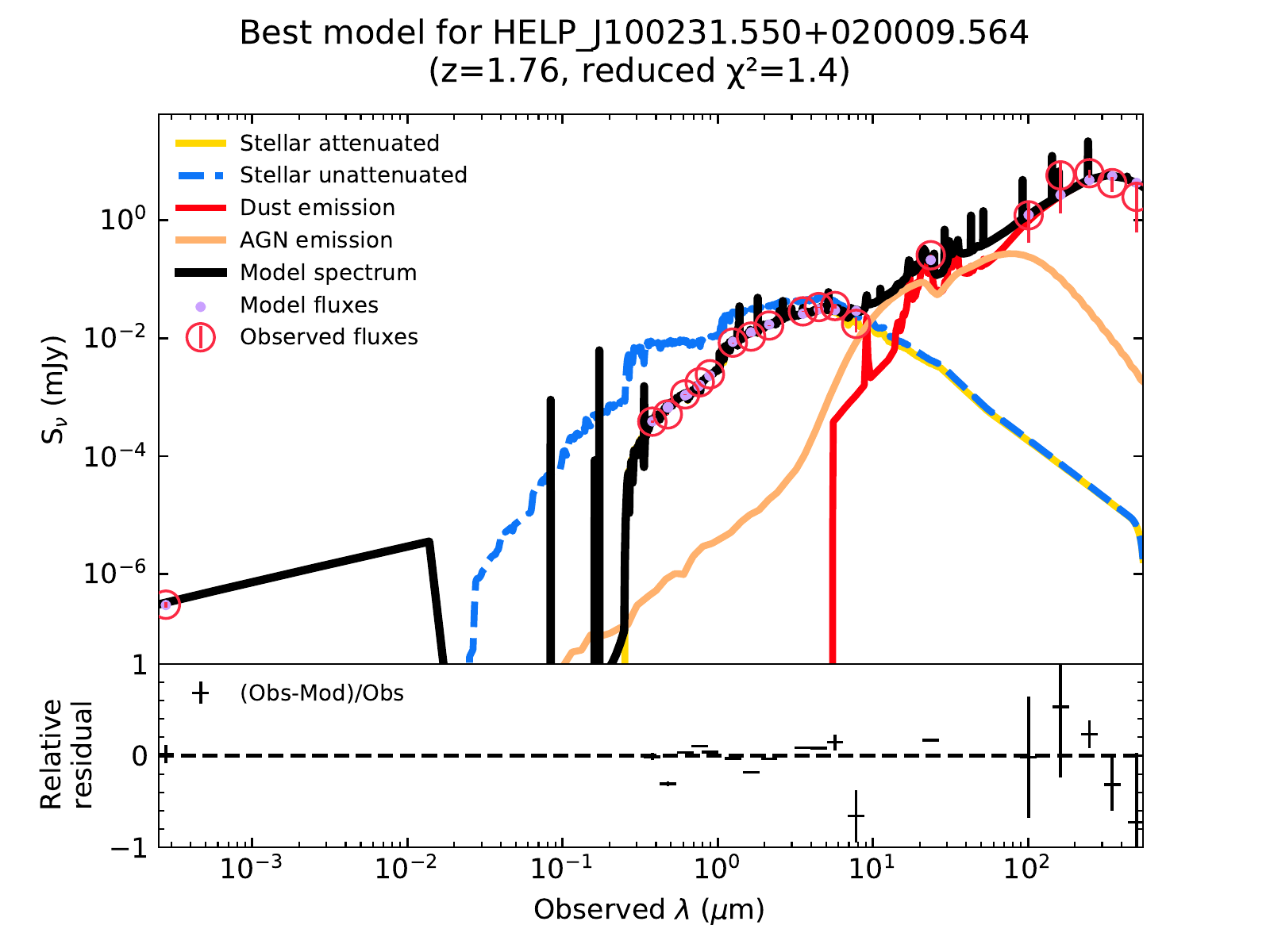}
  \label{}
  \end{subfigure}
\begin{subfigure}{.45\textwidth}
  \centering
  \includegraphics[width=1.\linewidth, height=7cm]{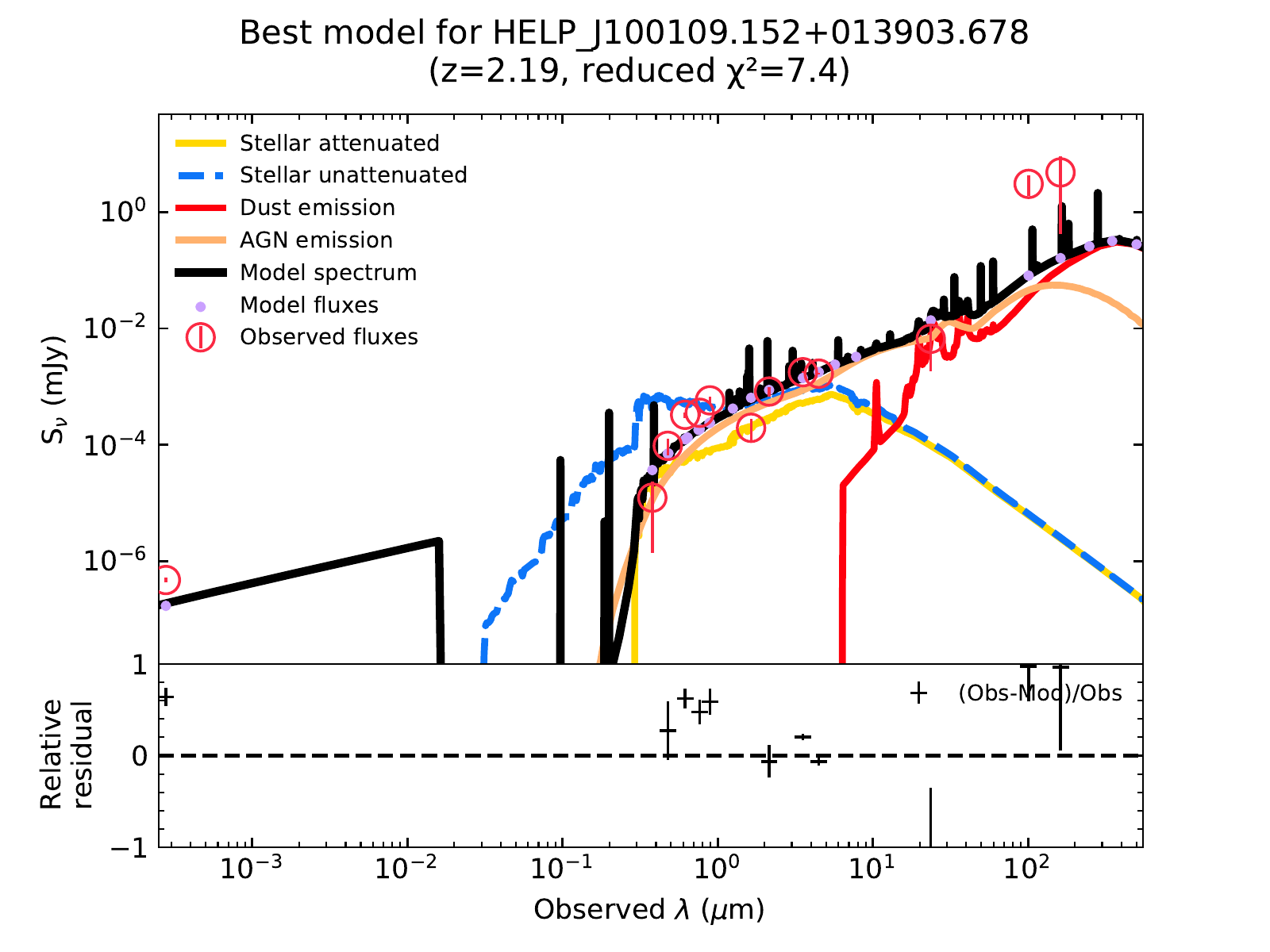}
  \label{}
\end{subfigure}
\begin{subfigure}{.45\textwidth}
  \centering
  \includegraphics[width=1.\linewidth, height=7cm]{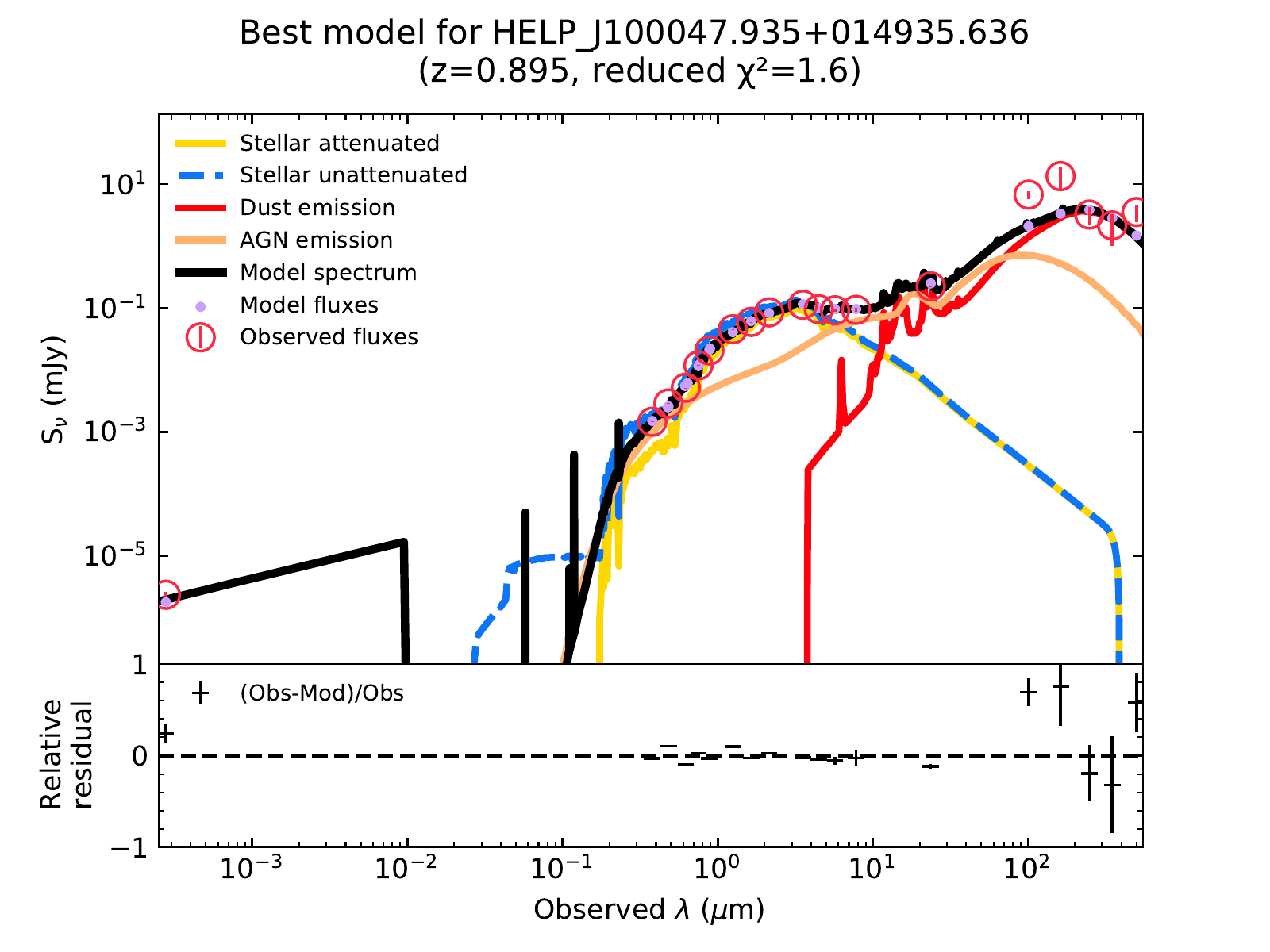}
  \label{}
  \end{subfigure}
  \caption{Examples of SEDs of AGN that meet our quality criteria (top panel) and do not satisfy (at least one) of our quality requirements (bottom panel). The AGN SED presented in the left, bottom panel is rejected from our analysis, due to its large $\chi^2_{red}$ value, while that on the right, bottom panel is rejected because it does not meet the $\rm \frac{1}{5}\leq \frac{SFR_{best}}{SFR_{bayes}} \leq 5$ criterion (see text for more details).}
  \label{fig_sed_agn}
\end{figure*}

\begin{figure*}
\centering
\begin{subfigure}{.45\textwidth}
  \centering
  \includegraphics[width=1.\linewidth, height=7cm]{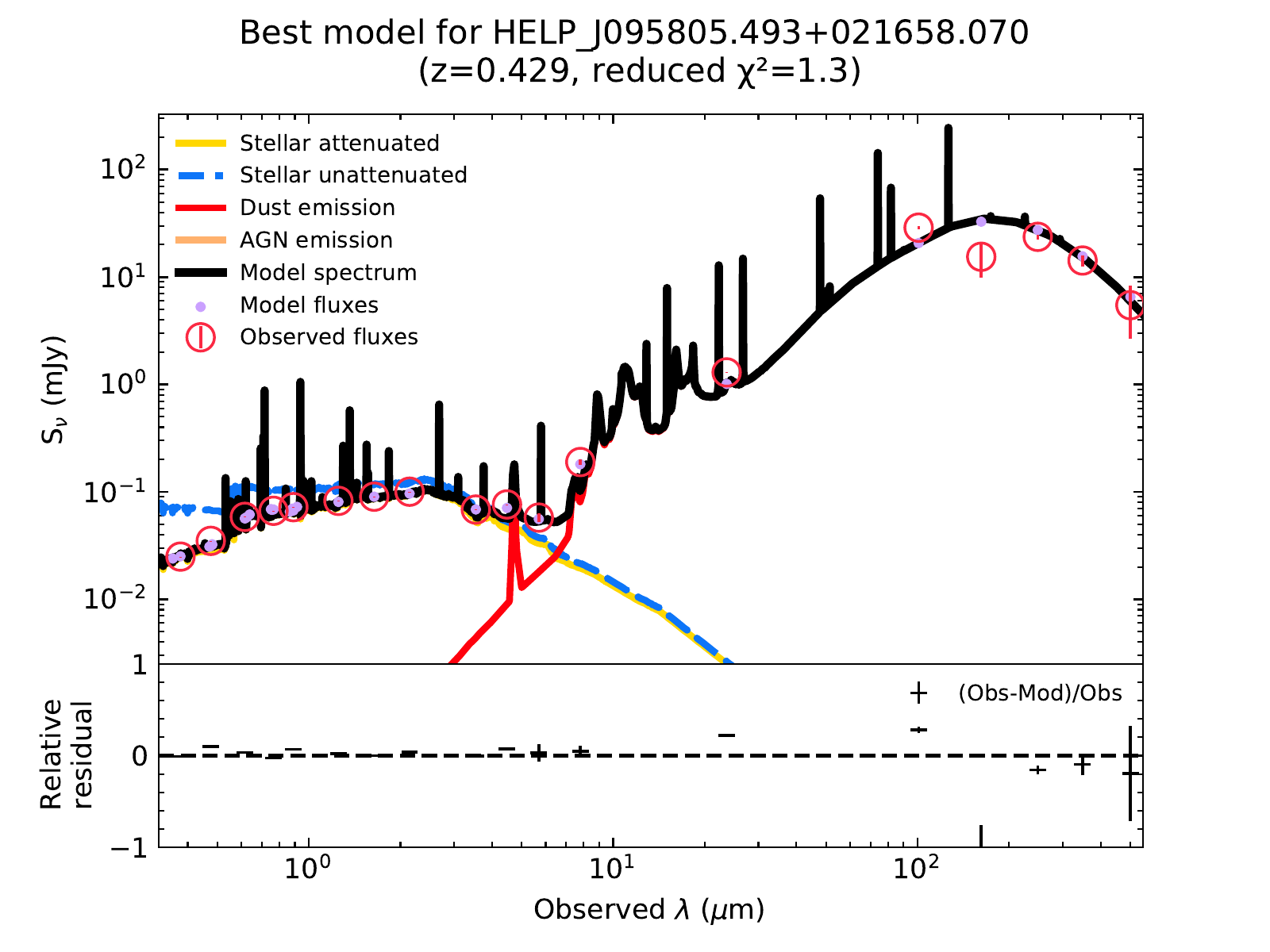}
  \label{}
\end{subfigure}
\begin{subfigure}{.45\textwidth}
  \centering
  \includegraphics[width=1.\linewidth, height=7cm]{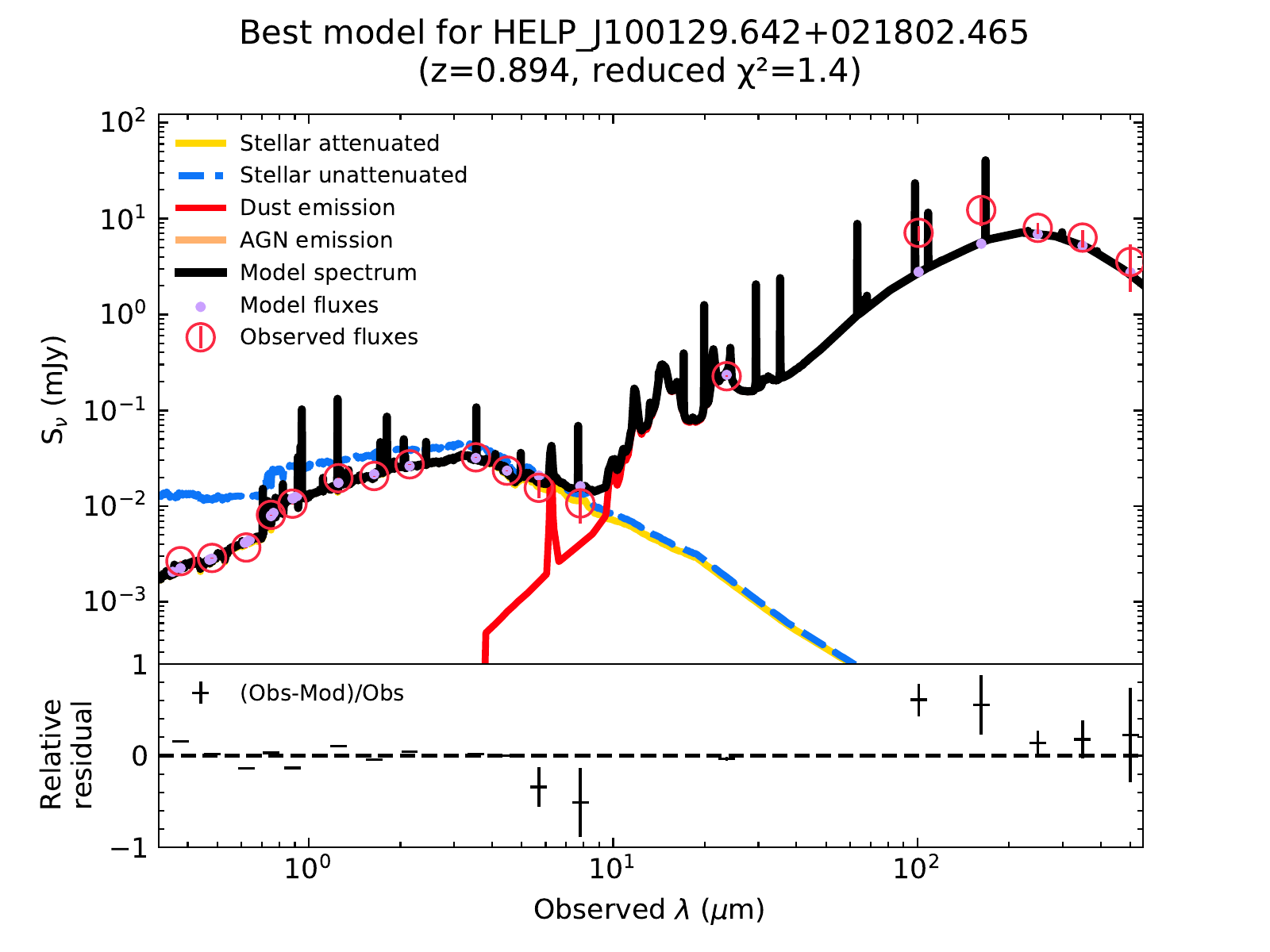}
  \label{}
  \end{subfigure}
\begin{subfigure}{.45\textwidth}
  \centering
  \includegraphics[width=1.\linewidth, height=7cm]{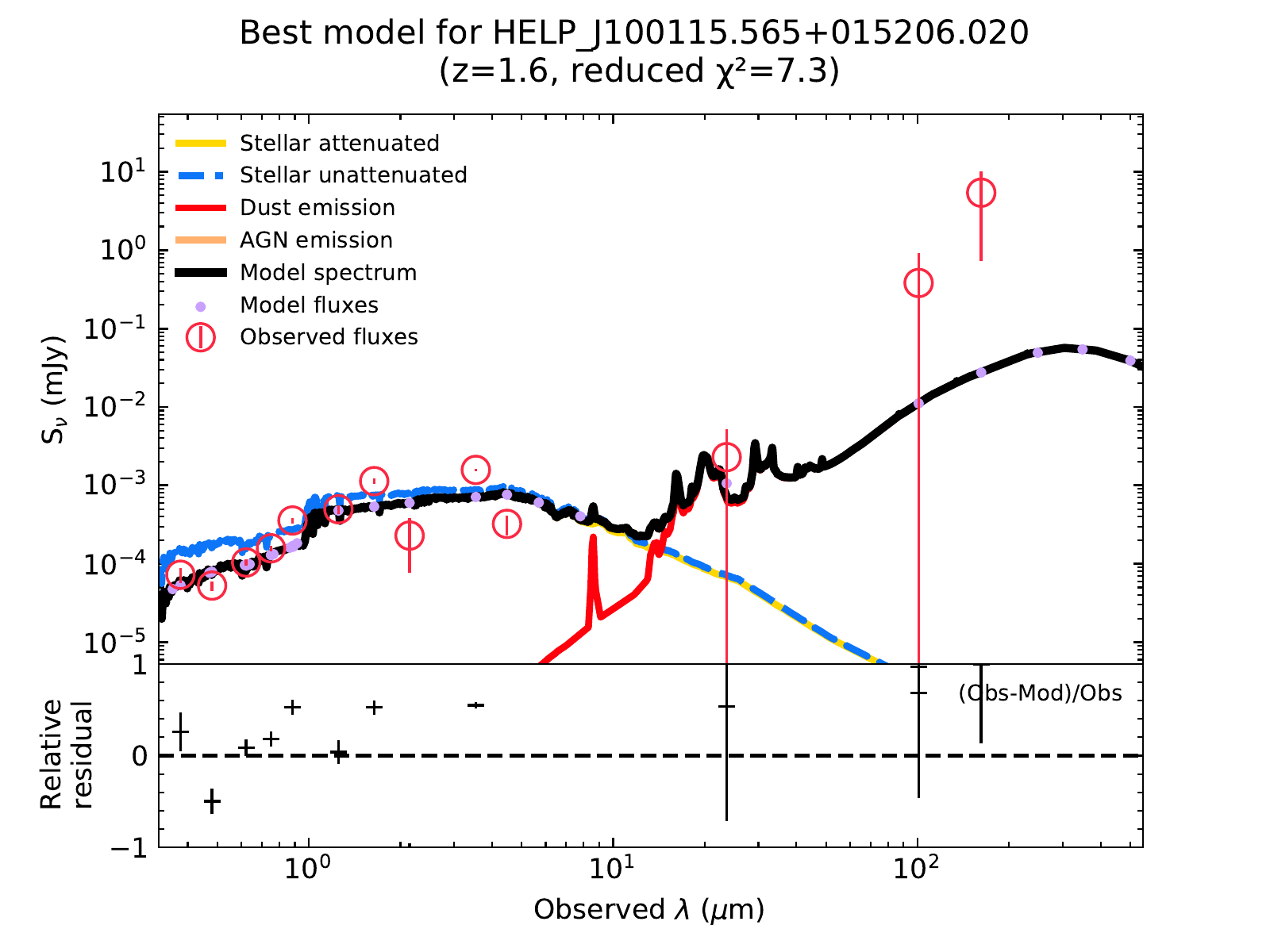}
  \label{}
\end{subfigure}
\begin{subfigure}{.45\textwidth}
  \centering
  \includegraphics[width=1.\linewidth, height=7cm]{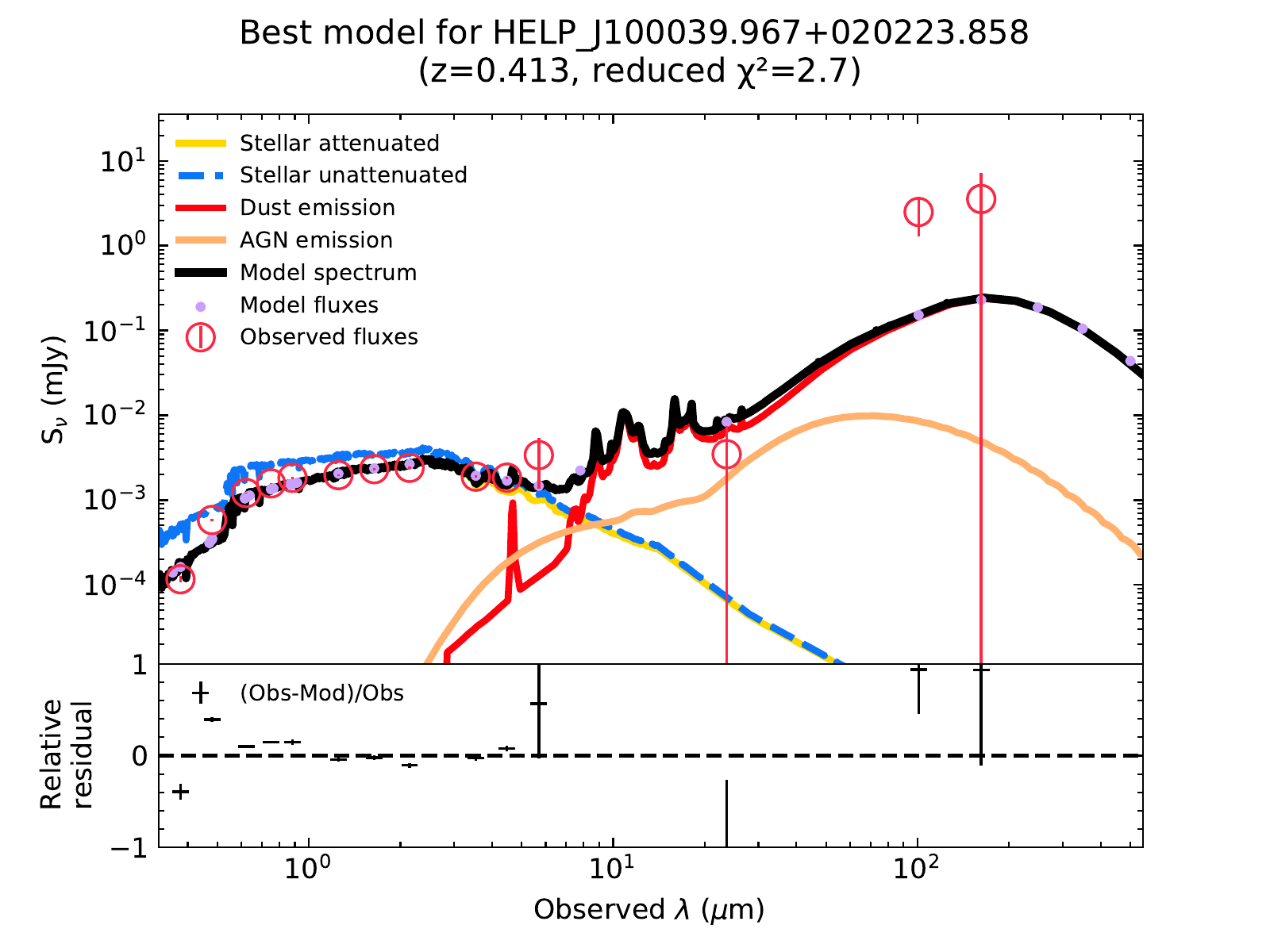}
  \label{}
  \end{subfigure}
  \caption{Examples of SEDs of sources in the reference galaxy catalogue that meet our quality criteria (top panel) and do not satisfy (at least one) of our quality requirements (bottom panel). The source with the SED presented in the left, bottom panel is rejected from our analysis, due to its large $\chi^2_{red}$ value, while that on the right, bottom panel is rejected because it does not meet the $\rm \frac{1}{5}\leq \frac{SFR_{best}}{SFR_{bayes}} \leq 5$ criterion (see text for more details).}
  \label{fig_sed_gals}
\end{figure*}

\begin{figure}
\centering
\begin{subfigure}{.5\textwidth}
  \centering
  \includegraphics[width=1.\linewidth, height=7cm]{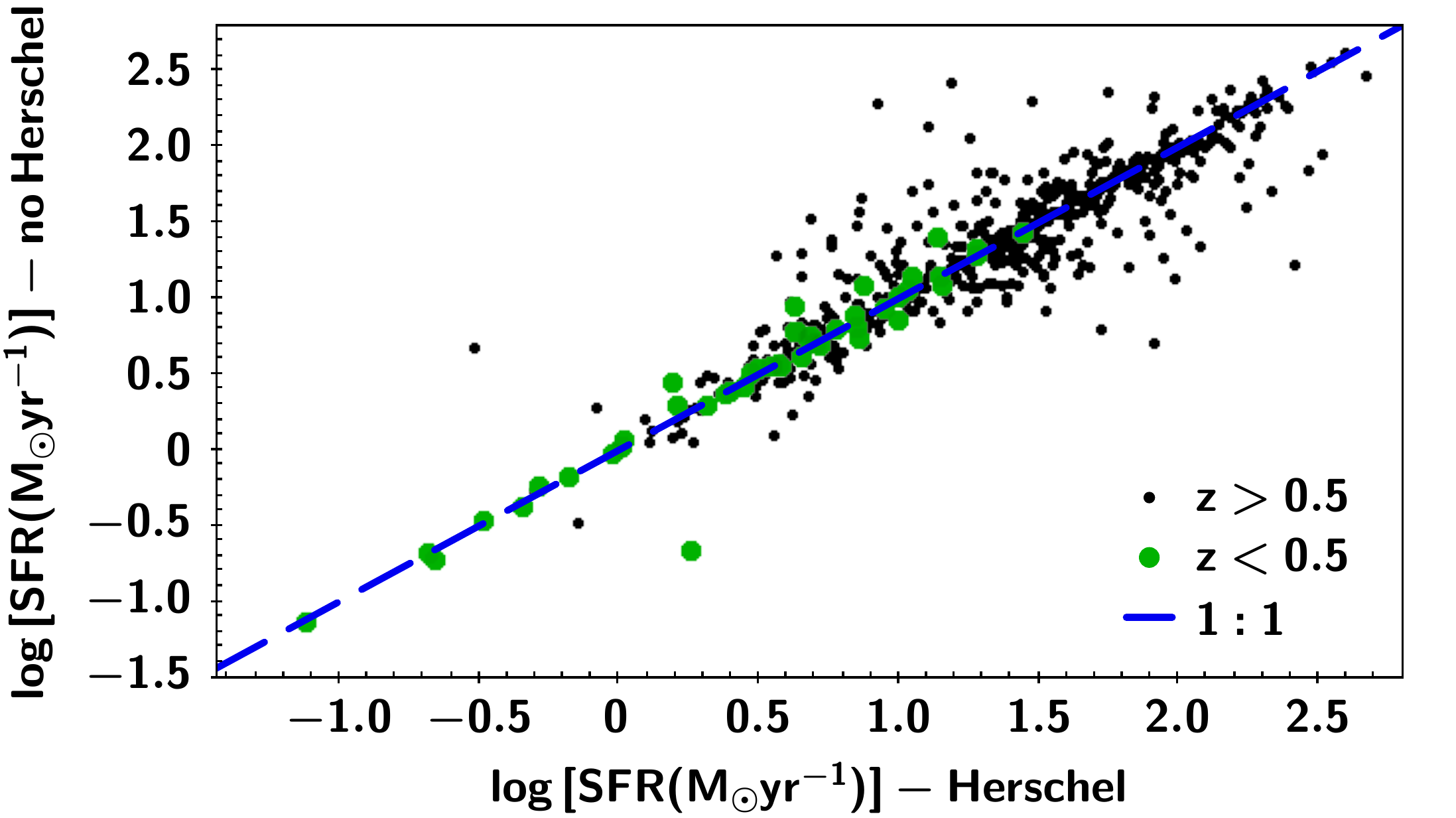}
  \label{}
\end{subfigure}
\begin{subfigure}{.5\textwidth}
  \centering
  \includegraphics[width=1.\linewidth, height=7cm]{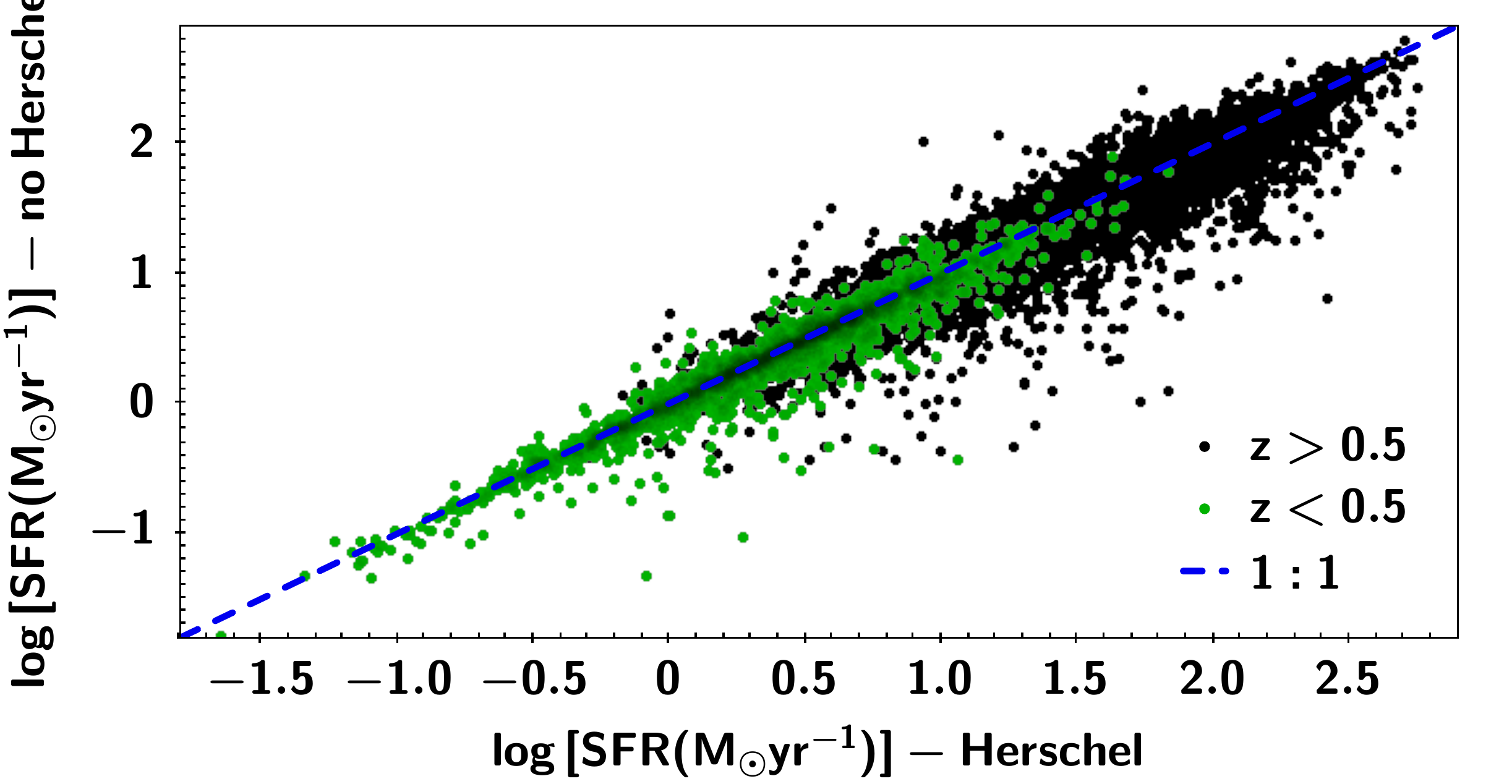}
  \label{}
  \end{subfigure}
  \caption{Comparison of SFR measurements with and without {\it{Herschel}}. Top panel: SFR calculations for 742 X-ray AGN with available far-IR photometry. The blue dashed line presents the 1:1 relation. Sources at $\rm z>0.5$ are shown with black circles, while those at $\rm z<0.5$ are shown with green circles. The mean difference of the SFR calculations is 0.01 for the overall population (0.00 for sources at $\rm z<0.5$) and the dispersion is 0.25 (0.19 at $\rm z<0.5$). Bottom panel: SFR measurements with and without {\it{Herschel}} for sources in the galaxy reference catalogue. The mean difference of the SFR calculations is 0.05 for the overall population (0.04 for sources at $\rm z<0.5$) and the dispersion is 0.16 (0.13 at $\rm z<0.5$).}
  \label{fig_herschel_comp}
\end{figure}

\begin{figure}
\centering
  \includegraphics[width=0.9\linewidth, height=7.cm]{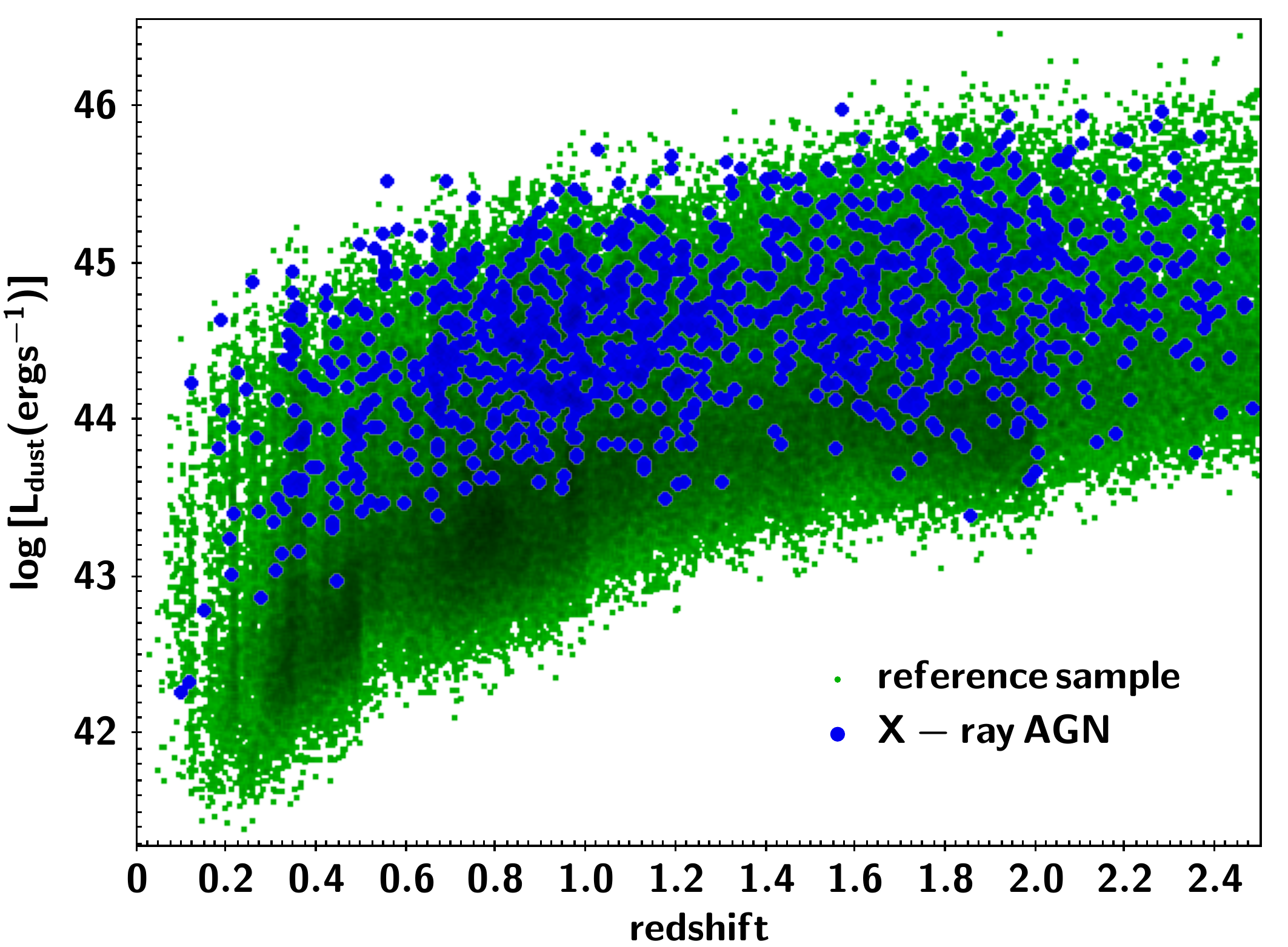}
  \caption{Dust luminosity, L$_{\rm dust}$, estimated by X-CIGALE, as a function of redshift, for sources in the reference catalogue (green points) and X-ray AGN (blue points). L$_{\rm dust}$ increases with redshift and, at all redshifts spanned by our datasets, AGN tend to reside in the more luminous systems.}
  \label{fig_lumz}
\end{figure} 

\section{Analysis}
\label{sec_analysis}

In this section, we describe the SED fitting analysis we applied to measure the galaxy properties. We, also, estimate the mass completeness of the data at different redshift intervals and describe the selection criteria we apply on the samples. We also examine the reliability of the X-CIGALE calculations.

\subsection{X-CIGALE}
\label{sec_cigale}

We use X-CIGALE \citep{Yang2020, Yang2022} to measure the properties of the galaxies in our datasets. X-CIGALE is a new branch of the CIGALE SED fitting code \citep{Boquien2019} that includes new features and improvements. Specifically, X-CIGALE has the ability to model the X-ray emission of galaxies. It can also model the extinction of the UV and optical emission in the poles of the AGN (polar dust). These improvements and their impact on the SED fitting of X-ray AGN have been examined in recent papers \citep{Yang2020, Mountrichas2021a, Mountrichas2021b, Buat2021}. 

In the SED fitting analysis, we use the same grid used in \cite{Mountrichas2021c}. This allows us to avoid any systematic effects introduced by using different modules and parameter space during the SED fitting process and facilitates a better comparison with the results presented in the Mountrichas et al. study. Here, we only summarize the modules included in the fitting process. 

A delayed star formation history (SFH) model with a function form $\rm SFR\propto t \times exp(-t/\tau$) is used to fit the galaxy component. The model includes a star formation burst in the form of an ongoing star formation no longer than 50\,Myr \citep{Buat2019}. The \cite{Bruzual_charlot2003} single stellar population template  is used to model the stellar emission. Stellar emission is attenuated following \cite{Charlot_Fall_2000}. The dust heated by stars is modelled following \cite{Dale2014}. The SKIRTOR template \citep{Stalevski2012, Stalevski2016} is used for the AGN emission. The same templates and parameter space are used for the X-ray and the galaxy reference sample, including the AGN template (SKIRTOR). The inclusion of the AGN module in the reference catalogue allows us to identify AGN with strong AGN component (see section \ref{sec_excl_agnfrac}). All free parameters used in the SED fitting process and their input values, are presented in Table \ref{table_cigale}.

As mentioned, X-CIGALE has the ability to model the X-ray emission of galaxies. In the SED fitting process, we use the observed X-ray luminosity in the 2$-$10\,keV band, provided by the \cite{Marchesi2016} catalogue. We confirm that using instead the intrinsic (absorption corrected) luminosities does not affect the host galaxy measurements. The photon index, $\Gamma$, is fixed to 1.4, i.e.,  the value assumed in the Marchesi et al. catalogue for their luminosity estimations.

\subsection{Quality and reliability examination of the fitting results}
\label{sec_bad_fits}


\begin{figure*}
\centering
\begin{subfigure}{.33\textwidth}
  \centering
  \includegraphics[width=1.0\linewidth, height=5.cm]{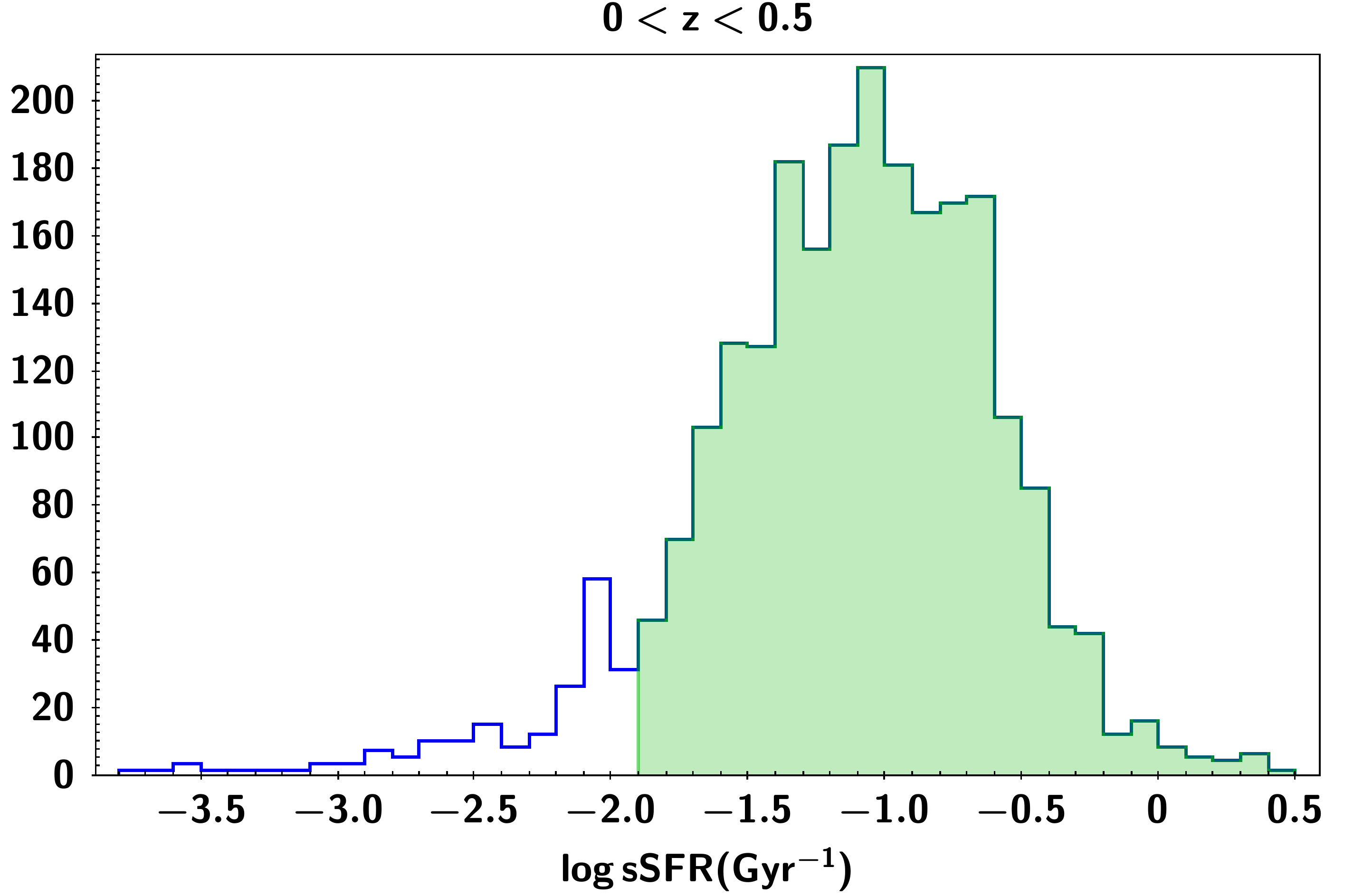}
  \label{}
\end{subfigure}
\begin{subfigure}{.33\textwidth}
  \centering
  \includegraphics[width=1.0\linewidth, height=5.cm]{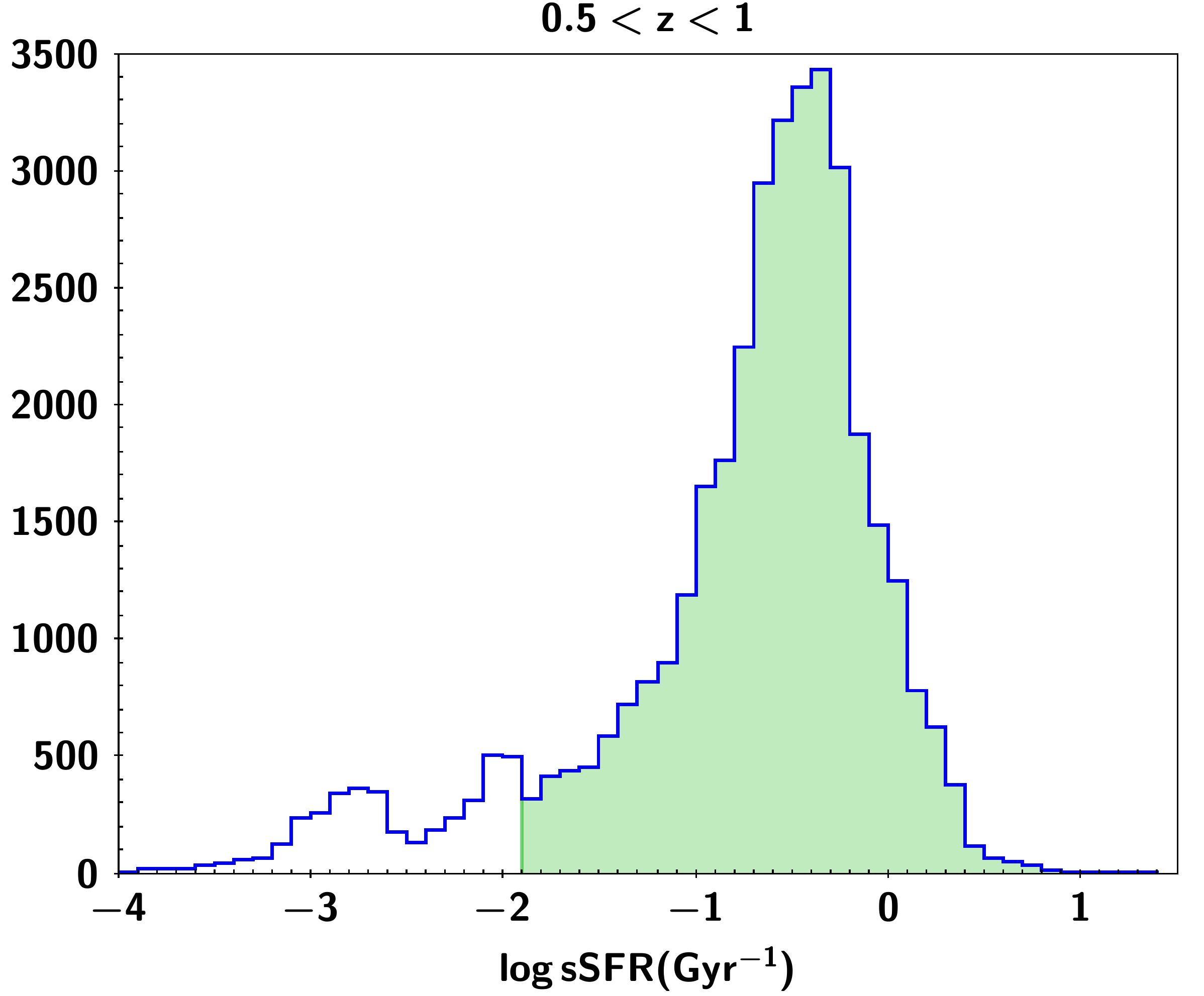}
  \label{}
\end{subfigure}
\begin{subfigure}{.33\textwidth}
  \centering
  \includegraphics[width=1.0\linewidth, height=5.cm]{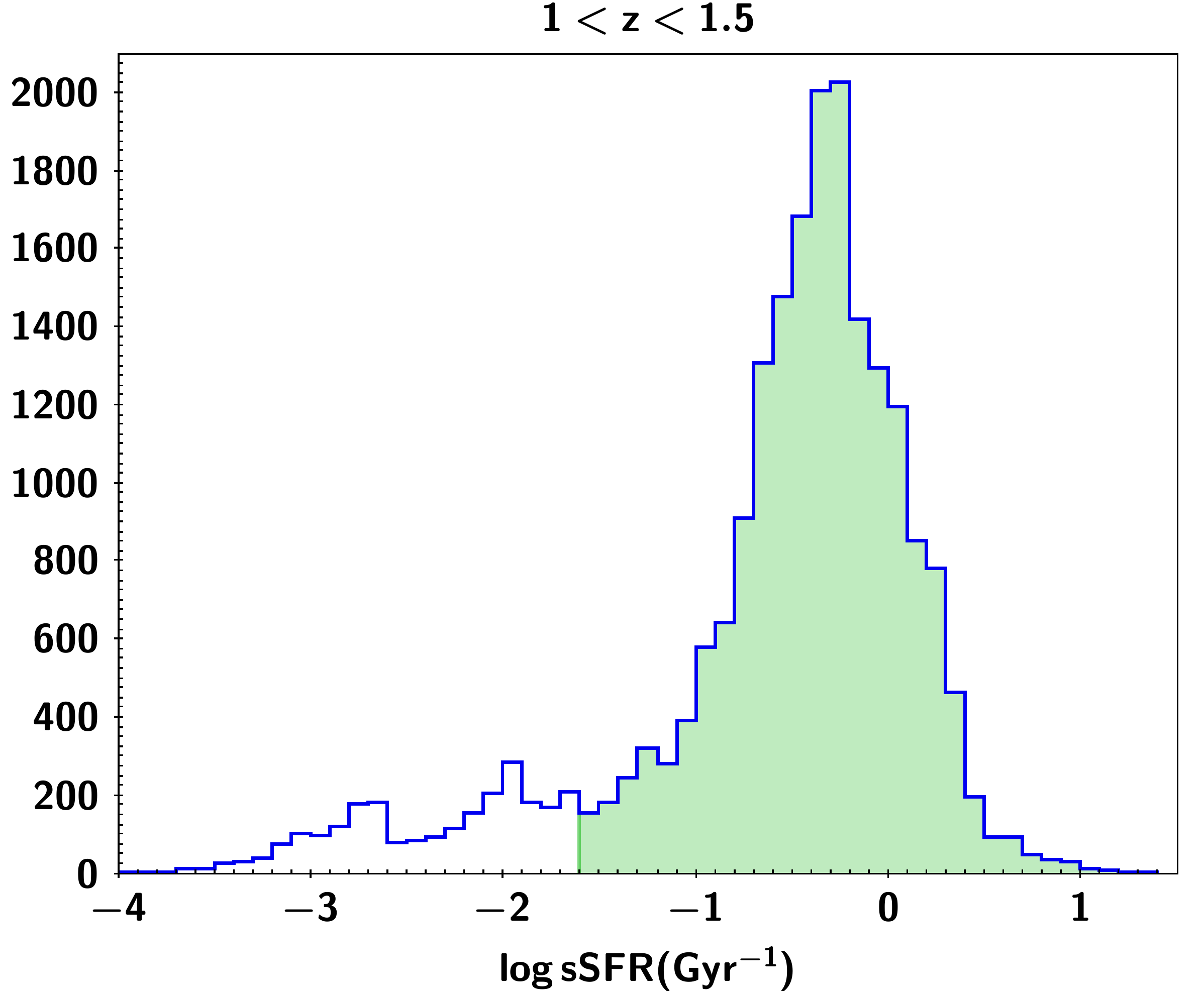}
  \label{}
\end{subfigure}
\begin{subfigure}{.33\textwidth}
  \centering
  \includegraphics[width=1.0\linewidth, height=5.cm]{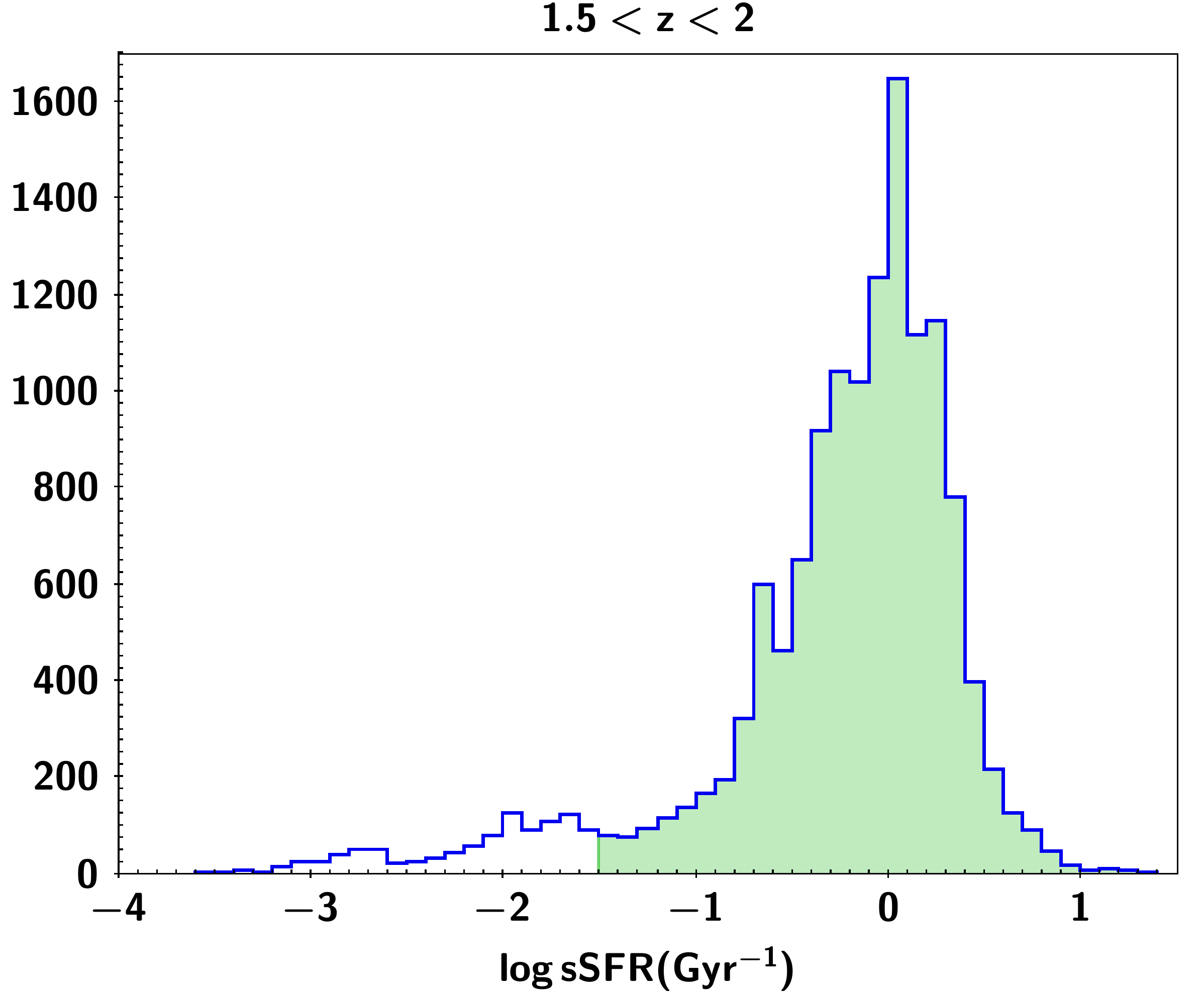}
  \label{}
\end{subfigure}
\begin{subfigure}{.33\textwidth}
  \centering
  \includegraphics[width=1.0\linewidth, height=5.cm]{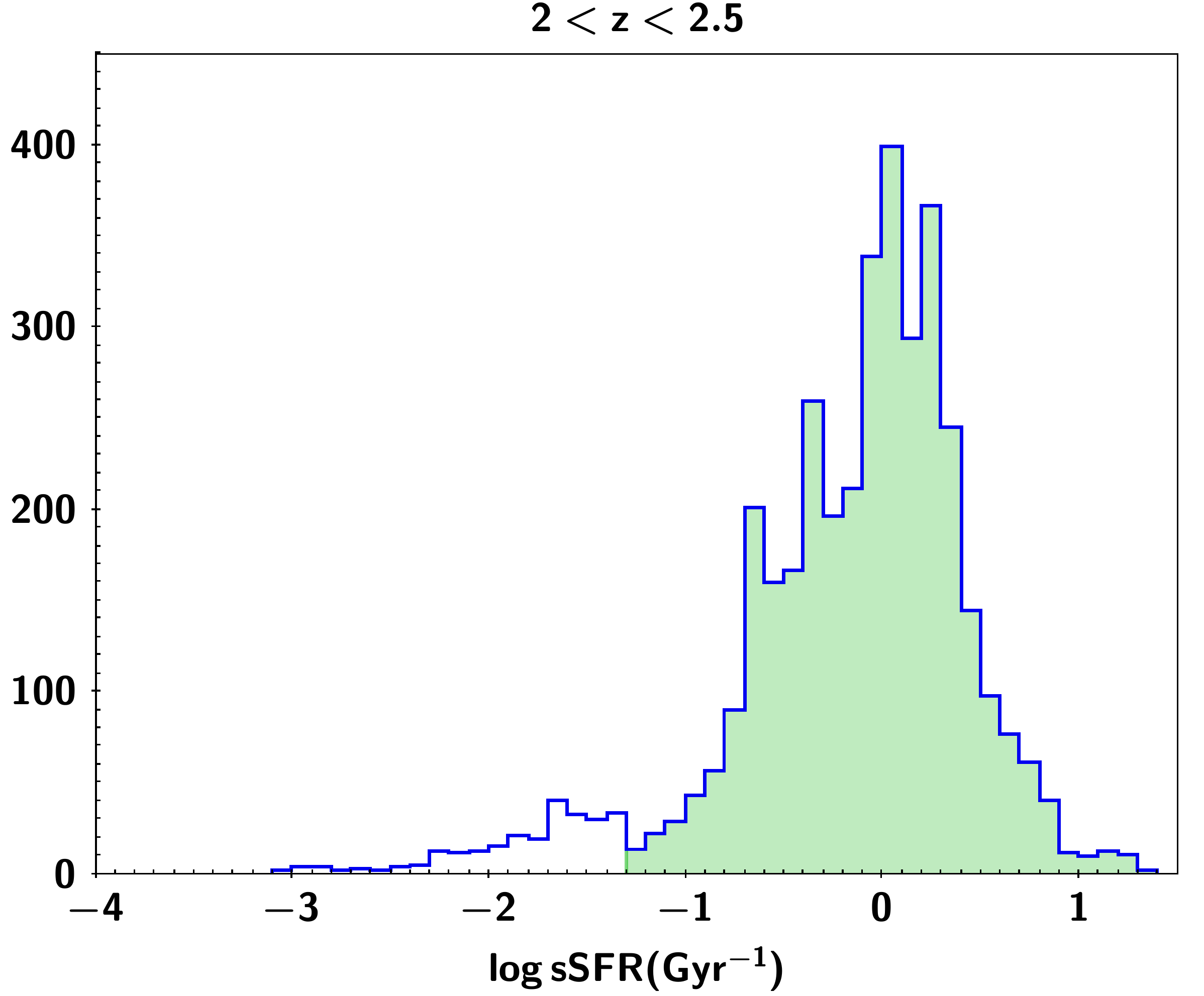}
  \label{}
\end{subfigure}
\caption{sSFR distributions of the reference galaxy sample, in five redshift intervals. Blue lines present the full distributions. Green areas are the sSFR distribution after applying the sSFR cut, which is defined based on the location of the second, lowest peak of each distribution.}
\label{fig_ssfr}
\end{figure*}

\subsubsection{Quality examination}
To keep in our analysis only sources with reliable host galaxy measurements, we exclude badly fitted SEDs. For that purpose, we consider only sources for which the reduced $\chi ^2$, $\chi ^2_r<5$. This value has been used in previous studies \citep[e.g.,][]{Masoura2018, Buat2021} and is based on visual inspection of the SEDs. This criterion is satisfied by 89\% and 92\% of the sources in the AGN and galaxy catalogues, respectively. 

To further exclude systems with non reliable measurements of the (host) galaxy properties, we apply the same method presented in \cite{Mountrichas2021c}, which has also been adopted in other studies \citep[e.g.,][]{Mountrichas2021b, Buat2021, Koutoulidis2021}. This method is based on the comparison of the value of the best model with the likelihood weighted mean value, calculated by X-CIGALE. Specifically, in our analysis, we only consider sources with $\rm \frac{1}{5}\leq \frac{SFR_{best}}{SFR_{bayes}} \leq 5$ and $\rm \frac{1}{5}\leq \frac{M_{*, best}}{M_{*, bayes}} \leq 5$, where SFR$\rm _{best}$, M$\rm _{*, best}$ are the best fit values of SFR and M$_*$, respectively and SFR$\rm _{bayes}$ and M$\rm _{*, bayes}$ are the Bayesian values, estimated by X-CIGALE. 1,236 and 177,245 X-ray AGN and galaxies in the reference catalogue fulfil these criteria, respectively. In Figures \ref{fig_sed_agn} and \ref{fig_sed_gals}, we present SEDs of sources that meet the aforementioned criteria (top panels) and of sources that are rejected from our analysis (bottom panels).

Furthermore, in our analysis, we take into account the uncertainties of the SFR and M$_*$ measurement that X-CIGALE provides. Each source is assigned two weights. One is based on the accuracy, $sigma$, of the SFR calculation ($sigma=\rm \frac{value}{error}$) and the other on the accuracy of the M$_*$ measurement.

\subsubsection{Reliability examination}

\cite{Mountrichas2021b,Mountrichas2021c}, used data from the XMM$-{\it XXL}$ and Bo$\rm \ddot{o}$tes fields and showed that lack of far-IR ({\it{Herschel}} photometry) does not affect the SFR calculations of X-CIGALE. We repeat the same check, using 742 AGN (60\% of the total X-ray sample) that have been detected by {\it{Herschel}}. For these sources, we perform SED fitting with and without {\it{Herschel}} bands, using the same parametric space. The results are shown in the top panel of Fig. \ref{fig_herschel_comp}. The mean difference, $\mu$, of the SFR measurements is 0.01 and the dispersion is $\sigma=0.25$. We repeat the same exercise for the sources in the reference galaxy catalogue. $\sim 20\%$ of them have Herschel detection. For this sample, the mean difference of the SFR calculations is 0.05 with $\sigma=0.16$ (bottom panel of Fig. \ref{fig_herschel_comp}). The dispersion of the measurements is independent of the AGN obscuration and galaxy extinction. We conclude that lack of far-IR photometry does not affect X-CIGALE's SFR calculations. 

UV photometry allows tracing the young stellar population of galaxies. In our SEDs, we do not include UV (GALEX) photometry, since this is available for less than 8\% of the sources. At $\rm z>0.5$, the $u$ band is redshifted to rest-frame wavelength $<2000\,\AA$, allowing observation of the emitted radiation from young stars. However, low redshift sources may have inaccurate SFR calculations when both UV and far-IR photometry are absent. Our sample includes 88 AGN that lie at $\rm z<0.5$. 70 of them have far-IR detection. These AGN are shown with green circles in Fig. \ref{fig_herschel_comp}. The mean difference in the SFR calculations is $\mu=0.00$ with dispersion $\sigma=0.19$. For sources at $\rm z<0.5$ in the galaxy reference catalogue, the mean SFR difference is 0.04 with $\sigma=0.13$. Therefore, the SFR calculation of sources at $\rm z<0.5$ that lack far-IR coverage is reliable and we include them in our analysis.

\begin{table*}
\caption{Number of X-ray AGN and sources in the reference galaxy catalogue, after applying the mass completeness limits at each redshift interval. In the parentheses we quote the number of sources, when we also exclude quiescent systems.}
\centering
\setlength{\tabcolsep}{1.mm}
\begin{tabular}{ccccccc}
 \hline
&total &$\rm z<0.5$&$\rm 0.5<z<1.0$ & $\rm 1.0<z<1.5$ & $\rm 1.5<z<2.0$ & $\rm 2.0<z<2.5$   \\
$\rm log (M_*/M_\odot)$ & &  ($> 8.60$) &  ($> 9.13$) & ($ > 9.44$) & ($> 9.69$) & ($ > 9.97$) \\
  \hline
X-ray catalogue &1,161 (852) & 88 (52) & 328 (246) & 272 (195) & 328 (243) & 145 (116)  \\
reference galaxy catalogue & 89,375 (80,475) &12,731 (11,438)  & 38,058 (34,104) & 21,124 (18,690) & 13,674 (12,697) & 3,788 (3,546)  \\
  \hline
\label{table_data}
\end{tabular}
\end{table*}

\subsection{Identification of non-X-ray AGN systems}
\label{sec_excl_agnfrac}

As mentioned in section \ref{sec_cigale}, in the SED fitting analysis we include an AGN template (SKIRTOR) when we fit the galaxy reference catalogue. This enables us to identify systems with a strong AGN component, by measuring the AGN fraction, $\rm frac_{AGN}$, parameter of X-CIGALE. $\rm frac_{AGN}$ is defined as the ratio of the AGN IR emission to the total IR emission of the galaxy ($1-1000\,\mu m$). We exclude from the reference catalogue, galaxies with $\rm frac_{AGN}>0.2$. For comparison, the median $\rm frac_{AGN}$ of the X-ray sample is 0.36. $\sim 33\%$ of sources in the galaxy catalogue are rejected. The fraction raises from lower ($\sim 13\%$ at $\rm z<1.0$) to higher (50-60\% at $\rm z>1.5$) redshifts. This increase is partially due to the fact that AGN tend to reside in more luminous systems that are found at higher redshifts (Fig. \ref{fig_lumz}). Moreover, recent observational studies have found that the AGN duty-cycle, i.e., the probability of a galaxy hosting an AGN within a given X-ray luminosity interval, increases with redshift \citep[e.g.][]{Georgakakis2017, Aird2018}. 124,815 galaxies are left in the reference galaxy sample. 

We note, that if we include all sources of the reference catalogue in our analysis, regardless of their AGN fraction value, the median SFR of galaxies is $\rm log\,[SFR (M_\odot yr^{-1})]=0.54$ (median $\rm log\,[M_*(M_\odot)]=10.0$) compared to $\rm log\,[SFR (M_\odot yr^{-1})]=0.59$ (median $\rm log\,[M_*(M_\odot)]=9.95$) when we exclude sources with $\rm frac_{AGN}>0.2$. Including these sources in our analysis, would increase the SFR$_{norm}$ values, presented in the next section, by on average $\approx 17\%$. In a similar manner, if we exclude from the reference catalogue sources with $\rm frac_{AGN}>0.1$, would decrease the SFR$_{norm}$ values by $\sim 5\%$ compared to those calculated when excluding galaxies with $\rm frac_{AGN}>0.2$. These fluctuations of the SFR$_{norm}$ caused by the choice of the $\rm frac_{AGN}$ values are within the statistical uncertainties of our measurements and do not affect our overall results and conclusions.

\begin{figure}
\centering
  \includegraphics[width=1.\linewidth, height=8.cm]{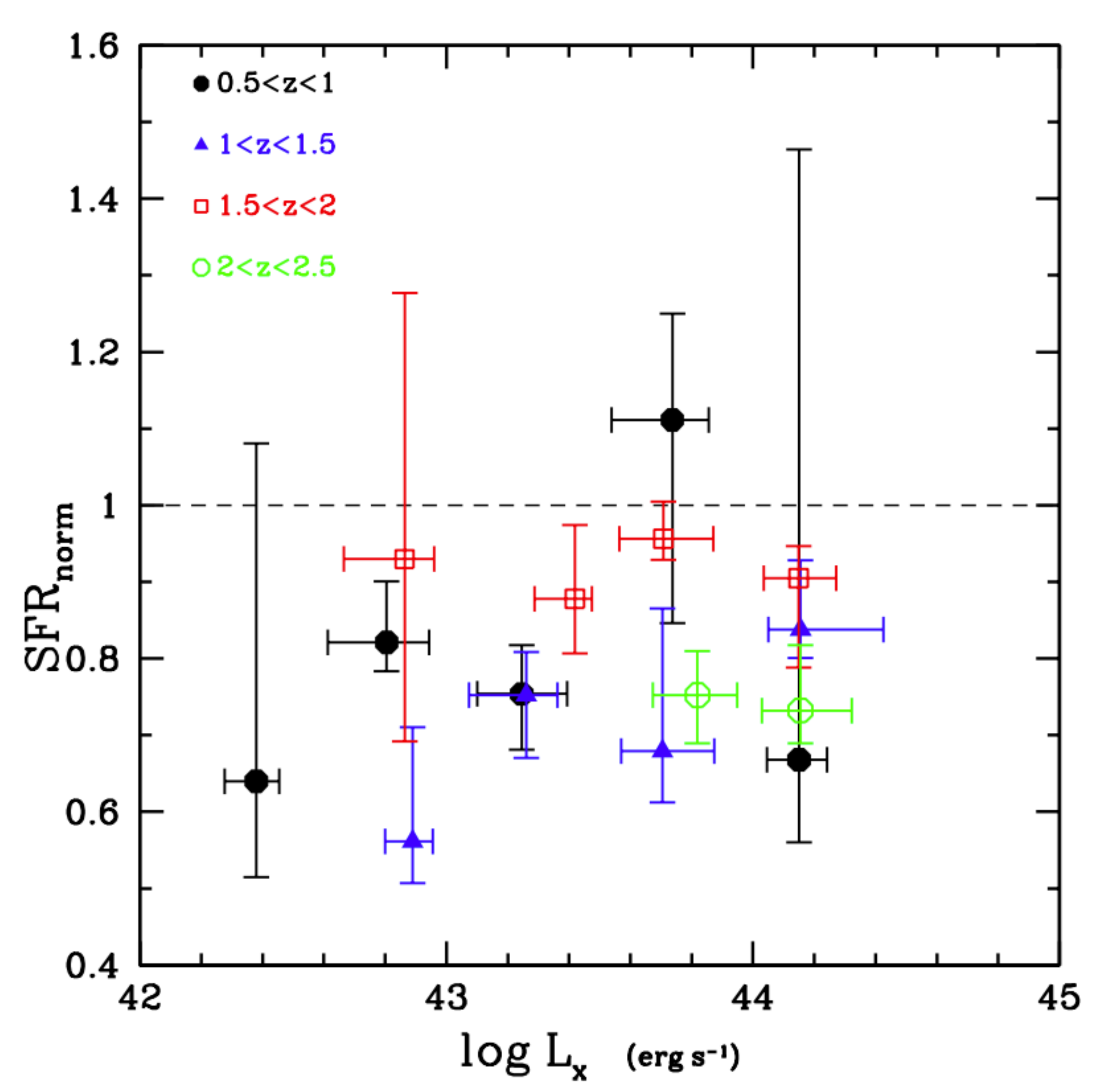}
  \caption{SFR$_{norm}$ vs. X-ray luminosity. SFR$_{norm}$ and L$_X$ are the median values of our binned measurements, in bins of L$_X$, with 0.5\,dex width. Errors are calculated using bootstrap resampling, by performing 1000 resamplings with replacement, at each bin. Results are colour coded based on the redshift interval. At all redshifts, SFR$_{norm}$ does not appear to evolve with L$_X$, at least up to luminosities $\rm 2\times 10^{44} erg/s$, spanned by the X-ray sample. SFR of X-ray AGN appears lower compared to that of star forming galaxies (i.e., SFR$_{norm}<1$). Although, for some of the bins, this is not statistical significant, it is consistent throughout the luminosities probed by the dataset.}
  \label{fig_sfrnorm_lx_redz}
\end{figure}


\begin{figure}
\centering
\begin{subfigure}{.5\textwidth}
  \centering
  \includegraphics[width=1.\linewidth, height=7cm]{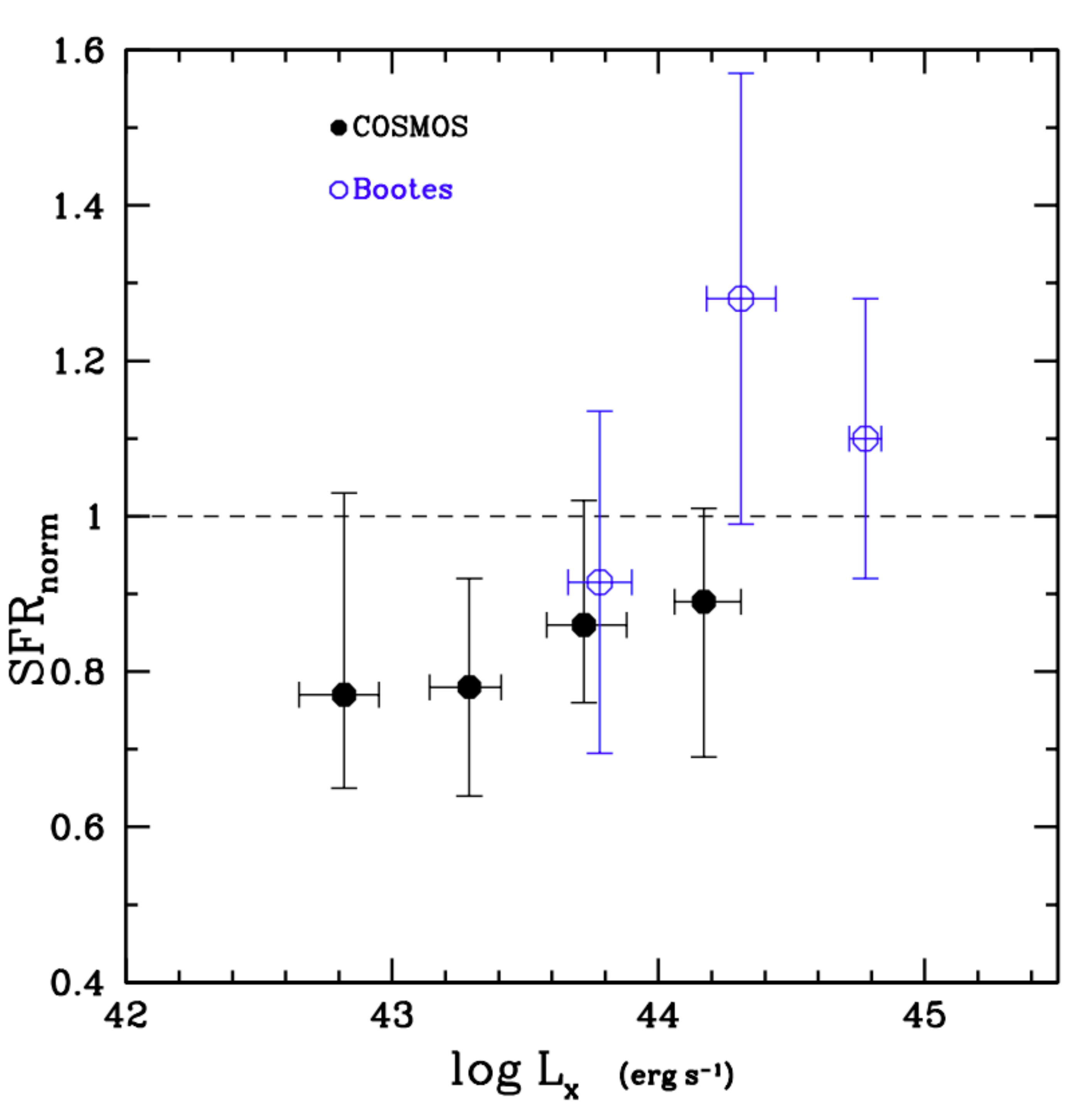}
  \label{}
\end{subfigure}
  \caption{SFR$_{norm}$ and L$_X$ are the mean values of the measurements presented in Fig. \ref{fig_sfrnorm_lx_redz}, grouped in L$_X$ bins of 0.5\,dex, at all redshifts combined, and weighted based on the number of sources in each bin shown in Fig. \ref{fig_sfrnorm_lx_redz}. Errors represent the standard deviation of SFR$_{norm}$ and L$_X$ in each bin. SFR$_{norm}$ is systematically lower than one (below the dashed line), which means that at these luminosities the SFR of X-ray AGN is lower compared to that of MS star forming galaxies. In the Bo$\rm \ddot{o}$tes field (blue circles), X-ray sources span higher luminosities than their counterparts in COSMOS. In the latter case, SFR$_{norm}$ values indicate the the SFR of AGN is at least equal (if not higher) than that of non-AGN systems.}
  \label{fig_sfrnorm_lx_ave_bootes}
\end{figure}

\subsection{Mass completeness}
\label{sec_mass_completeness}

In our analysis, we study the SFR of X-ray AGN as a function of X-ray luminosity, in five redshift bins, up to redshift 2.5. To avoid possible biases introduced by the different mass completeness of our datasets at different redshift intervals, we calculate the mass completeness at each redshift bin. For that, we follow the method described in \cite{Pozzetti2010} and apply it on the galaxy reference sample due to its significantly larger size. The same method has been followed in similar studies \citep{Florez2020, Mountrichas2021c} and has also been used to estimate the stellar mass completeness of the COSMOS2015 galaxy catalogue, presented in \cite{Laigle2016}.   

First, we estimate the limiting stellar mass, M$_{*,lim}$, of each galaxy, using the following expression:

\begin{equation}
\rm log M_{*,lim} = log M_*+0.4(m-m_{lim}),
\end{equation}
where M$_*$ is the stellar mass of each source, measured by X-CIGALE, m is the AB magnitude of the source and m$_{\rm lim}$ is the AB magnitude limit of the survey. This equation essentially calculates the mass the galaxy would have if its apparent magnitude was equal to the limiting magnitude of the survey for a specific photometric band. Then, we use the $\rm log M_{*,lim}$ of the 20\% of the faintest galaxies at each redshift bin. The minimum stellar mass at each redshift interval for which our sample is complete is the 95th percentile of $\rm log M_{*,lim}$, of the $20\%$ faintest galaxies in each redshift bin.

We follow \cite{Laigle2016} and use $\rm K_s$ as the limiting band of the samples with value, $\rm K_s=24.7$ \citep[see Table 1 in][]{Laigle2016}. We find that the stellar mass completeness of our galaxy reference catalogue is $\rm log\,[M_{*,95\%lim}(M_\odot)]= 8.60, 9.13, 9.44, 9.69$ and 9.97 at $\rm z<0.5, 0.5<z<1.0$, $\rm 1.0<z<1.5$, $\rm 1.5<z<2.0$ and $\rm 2.0<z<2.5$, respectively. These values agree with those presented in Table 6 of \cite{Laigle2016}, although the redshift intervals used are slightly different in the two studies.

\subsection{Identification of quiescent systems}
\label{sec_MS}

The main purpose of this work is to study the position of X-ray AGN relative to the MS, as a function of X-ray luminosity and redshift. Towards this end, we follow previous studies \citep[e.g.][]{Mullaney2015, Masoura2018, Bernhard2019, Masoura2021} and estimate the SFR$_{norm}$ parameter. SFR$_{norm}$ is defined as the ratio of the SFR of AGN to the SFR of MS galaxies with similar stellar mass and redshift. However, we want to avoid systematic effects that may have affected previous works that estimated SFR$_{norm}$ using analytical expressions from the literature to parametrise the SFR of MS galaxies \citep[e.g., equation 9 of][]{Schreiber2015}. In our analysis, SFR has been calculated in an homogeneous manner, both for the X-ray and the reference catalogues, with the same wavelength coverage which enables us to minimise systematic effects. Thus, we follow the process described in \cite{Mountrichas2021c} to define our own star forming main sequence (SFMS). The goal is not to make a strict definition of the MS, but to exclude in a uniform manner the majority of quiescent systems from our data.  

For this exercise, we use the galaxy reference catalogue due to its large size. The outcome of this process is then used in both the galaxy and the X-ray AGN samples to exclude quiescent sources. First, we estimate the sSFR ($\rm sSFR=\frac{SFR}{M_*}$)  of each source. Fig. \ref{fig_ssfr} presents the distributions of sSFR in each redshift interval.  We see that each distribution presents a long tail or a lower second peak at low sSFR values. We consider that these peaks are populated by quiescent systems. Table \ref{table_data} presents the number of sources in each redshift interval, after excluding quiescent systems. Based on our results, $\sim$ 10\% of the galaxies in the reference catalogue and $\sim$ 25\% of AGN are quiescent. This result indicates that a larger fraction of X-ray AGN reside in quiescent systems compared to non-AGN galaxies, up to redshift $\rm z<2.5$. Moreover, this fraction raises for the X-ray sources from 20\% at $\rm z>2$ to 40\% at $\rm z<0.5$. This is in agreement with previous studies that found an increased fraction of AGN hosted by quiescent galaxies at low redshifts compared to higher redshifts \citep{Shimizu2015, Koutoulidis2021}.


\section{Results}
\label{sec_lx_sfr}

In this section, we compare the SFR of X-ray AGN with the SFR of non-AGN systems, at different luminosity and redshift intervals. Next, we examine whether the results of this comparison (also) depend on the stellar mass of the host galaxy.

\subsection{SFR of X-ray AGN relative to star forming galaxies, as a function of luminosity and redshift} 
\label{sfrnorm_lx_z}

To compare the SFR of X-ray AGN with that of star forming galaxies, we use the SFR$_{norm}$ parameter. To estimate SFR$_{norm}$, we follow the method presented in \cite{Mountrichas2021c}. Specifically, the SFR of each X-ray source is divided by the SFR of galaxies from the reference catalogue that have stellar mass that differs $\pm 0.1$\,dex from the stellar mass of the AGN and lies within $\pm 0.075 \times (1 + \rm z)$ from the X-ray source. The median value of these ratios is used as the SFR$_{norm}$ of each AGN. In these calculations, each source is weighted based on the uncertainty on the SFR and M$_*$ (see Sect. \ref{sec_bad_fits}). We only keep X-ray sources for which the SFR$_{norm}$ has been estimated using at least 30 galaxies from the reference catalogue. This limit lowers to at least 20 galaxies for the highest redshift bin ($\rm 2.0<z<2.5$) and goes down to ten galaxies when we consider only the most massive systems ($\rm 11.5< log\,[M_*(M_\odot)] < 12.0$, sect. \ref{sfrnorm_lx_mstar}).

Fig. \ref{fig_sfrnorm_lx_redz}, presents SFR$_{norm}$ as a function of X-ray luminosity, for different redshift intervals. Both for the X-ray sample and the galaxy reference catalogue, quiescent systems have been excluded. In this exercise, we do not present results for the lowest redshift bin ($\rm z<0.5$) due to the limited number of X-ray sources. Median values are presented. Errors are calculated using bootstrap resampling \citep[e.g.][]{Loh2008}, by performing 1000 resamplings with replacement at each bin. Results are grouped in L$_x$ bins of 0.5\,dex. Bins with fewer than 25 AGN are not presented, throughout this work. The results do not show dependence of the SFR$_{norm}-$L$_X$ relation with redshift, in agreement with previous studies \citep{Mullaney2015, Mountrichas2021c}. However, we notice that the SFR of X-ray AGN appears lower compared to that of star forming galaxies (i.e., SFR$_{norm}<1$). Although, for some of the bins, this is not statistically significant, it is consistent throughout the luminosities probed by our sample. 


Since there is no dependence on redshift, in Fig. \ref{fig_sfrnorm_lx_ave_bootes}, we present the weighted average SFR$_{norm}$ in bins of L$_X$, over the total redshift range. Mean values are weighted based on the number of sources included in each luminosity bin. Errors present the standard deviation of the measurements. We also plot the results from \cite{Mountrichas2021c} in the Bo$\rm \ddot{o}$tes field. The COSMOS X-ray sample spans about an order of magnitude lower luminosities compared to that in Bo$\rm \ddot{o}$tes. However, the latter reaches higher L$_X$. This allows us to draw a picture of SFR$_{norm}$ as a function of Lx, for about two orders of magnitudes in X-ray luminosity. In the overlapping L$_X$ regime the results from the two studies are consistent. Although the size of the X-ray samples, do not allow us to draw strong conclusions, at $\rm L_{X,2-10keV} < 10^{44}\,erg\,s^{-1}$, the SFR of AGN appears lower or equal to that of SFMS, ($\rm SFR_{AGN} \leq SFR_{SFMS}$), whereas at $\rm L_{X,2-10keV} > 2-3 \times 10^{44}\,erg\,s^{-1}$ there is an increase of SFR$_{norm}$, i.e.,  $\rm SFR_{AGN}\geq SFR_{SFMS}$.


Overall, our results show no evolution of the SFR$_{norm}-$L$_X$ relation with redshift. The SFR of AGN is equal or lower compared to that of MS galaxies at  $\rm L_{X,2-10keV} < 10^{44}\,erg\,s^{-1}$ and combining our measurements with those in the Bo$\rm \ddot{o}$tes field, there is a hint of an increase of SFR$_{norm}$ at $\rm L_{X,2-10keV} > 2-3 \times 10^{44}\,erg\,s^{-1}$.

\begin{table*}
\caption{Number of X-ray AGN and sources in the reference galaxy catalogue, in each stellar mass bin, after excluding quiescent systems and applying the mass completeness limits.}
\centering
\setlength{\tabcolsep}{1.5mm}
\begin{tabular}{ccccccc}
 \hline
$\rm log (M_*/M_\odot)$ &   total & $<10.0$ & $10.0-10.5$ & $10.5-11.0$ & $11.0-11.5$ & $11.5-12.0$ \\
  \hline
X-ray AGN & 852 & 15 & 81 & 273 & 396 & 87  \\
reference galaxy catalogue &80,475 & 41,558 & 20,192 & 12,511 & 5,581 & 633 \\
  \hline
\label{table_xray_mstar_bins}
\end{tabular}
\end{table*}

\subsection{SFR of X-ray AGN relative to star forming galaxies, as a function of luminosity and stellar mass}
\label{sfrnorm_lx_mstar}

Previous studies, using either the specific black hole accretion rate parameter \citep[$\lambda_{s,BHAR}$,][see also next section]{Georgakakis2017, Aird2018, Aird2019, Torbaniuk2021} or the specific X-ray luminosity \citep{Yang2018}, found that the stellar mass of the host galaxy, may affect how the AGN and galaxy properties are linked. \cite{Mountrichas2021c} found indications that the increase of SFR$_{norm}$ at $\rm L_{X,2-10keV} > 2-3 \times 10^{44}\,erg\,s^{-1}$ is more evident for AGN that live in galaxies that have stellar masses within a specific range ($\rm log\,[M_*(M_\odot)] \sim 11-11.5$) while the most massive ones do not present this feature in their SFR$_{norm}-$L$_X$ relation. Moreover, in the previous section, we found that the SFR$_{norm}$ measurements from the COSMOS X-ray sample are generally lower compared to those in the Bo$\rm \ddot{o}$tes field (Fig. \ref{fig_sfrnorm_lx_ave_bootes}), but also span lower X-ray luminosities. Another differentiating factor between the two samples, is that the X-ray sources in Bo$\rm \ddot{o}$tes are more massive than their counterparts in COSMOS, due to the higher mass completeness limits of the former sample. In COSMOS, at $\rm z<0.5$ the median stellar mass of galaxies that host X-ray AGN is $\rm log\,[M_*(M_\odot)] = 10.8$ and increases to $\rm log\,[M_*(M_\odot)] = 11.0-11.1$ at redshift bins higher than $\rm z>0.5$, i.e., they are about 0.5\,dex lower than the AGN hosts in Bo$\rm \ddot{o}$tes \citep[see Table 4 in][]{Mountrichas2021c}.

\begin{figure}
\centering
  \includegraphics[width=1.\linewidth, height=8.cm]{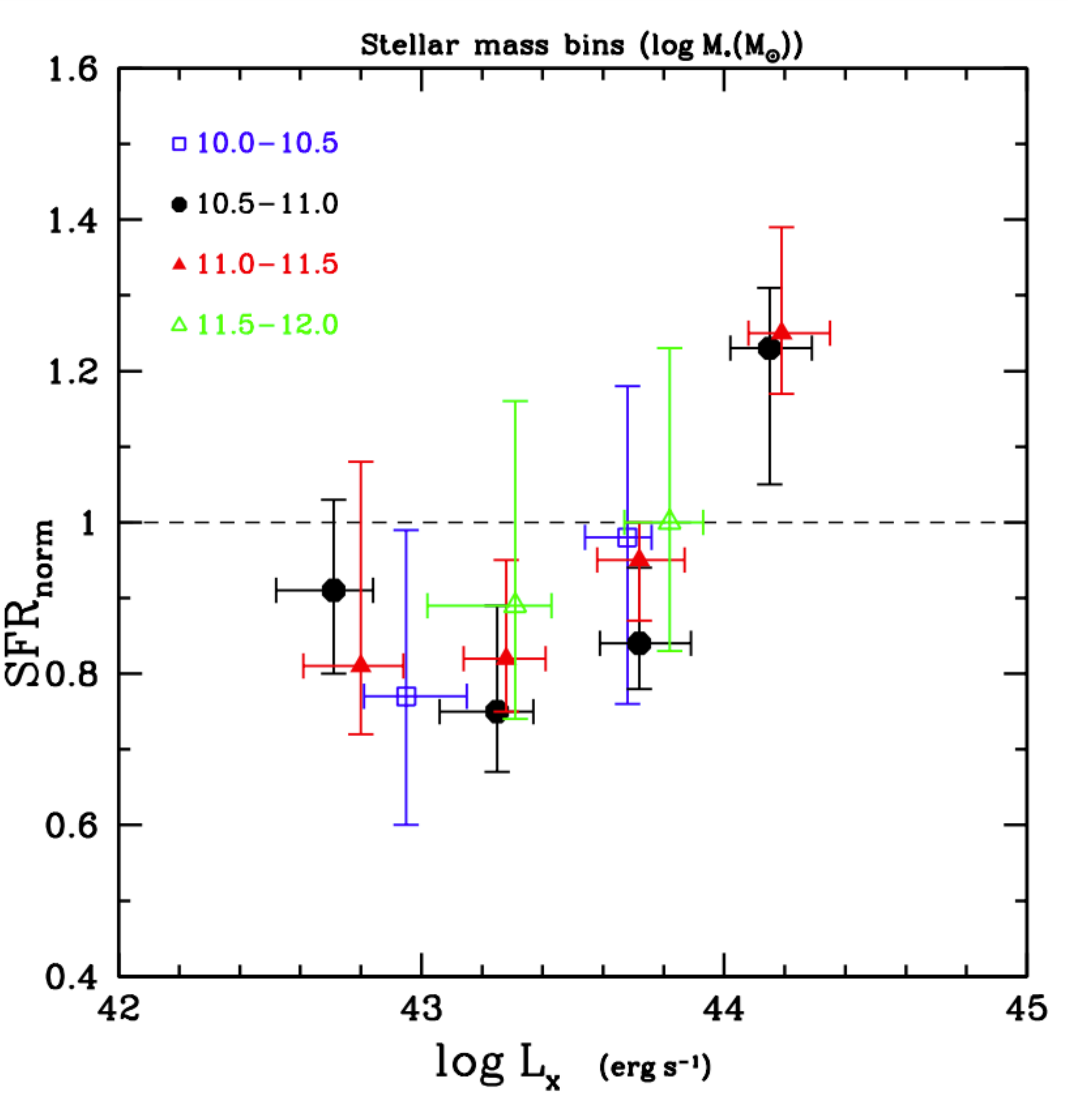}
  \caption{SFR$_{norm}$ vs. X-ray luminosity. Results are grouped in L$_X$ and stellar mass bins of 0.5\,dex width. Since there is no evolution of the SFR$_{norm}$-L$_X$ relation with redshift (Fig. \ref{fig_sfrnorm_lx_redz}), measurements are taken at all redshifts spanned by our sample ($\rm 0<z<2.5$). The SFR$_{norm}$-L$_X$ relation appears flat at $\rm L_{X,2-10keV}<10^{44}\,erg\,s^{-1}$. However, at $\rm L_{X,2-10keV}>10^{44}\,erg\,s^{-1}$, SFR$_{norm}$, increases with L$_X$.} 
  \label{fig_sfrnorm_lx_mstar}
\end{figure}

To disentangle the effect of stellar mass, we repeat the measurements presented in the top panel of Fig. \ref{fig_sfrnorm_lx_ave_bootes}, but we also group sources in stellar mass bins. In this exercise, we include X-ray AGN at $\rm z<0.5$ (52 sources, see Table \ref{table_data}). The number of X-ray AGN available at each stellar mass interval are shown in Table \ref{table_xray_mstar_bins}. Fig. \ref{fig_sfrnorm_lx_mstar} presents the results for four M$_*$ bins. At $\rm log\,[M_*(M_\odot)] < 10.0$ the number of AGN is too low to make a meaningful calculation. We notice that at $\rm L_{X,2-10keV} < 10^{44}\,erg\,s^{-1}$, regardless of the stellar mass of their host galaxy, AGN have $\rm SFR_{AGN} \leq SFR_{SFMS}$. At $\rm L_{X,2-10keV} > 10^{44}\,erg\,s^{-1}$, AGN with $\rm 10.5 < log\,[M_*(M_\odot)] < 11.0$ and $\rm 11.0 < log\,[M_*(M_\odot)] < 11.5$ show enhancement of their SFR compared to MS galaxies. Unfortunately, there are not enough AGN in the COSMOS field with $\rm log\,[M_*(M_\odot)] < 10.5$ and $\rm log\,[M_*(M_\odot)] > 11.5$ at $\rm L_{X,2-10keV} > 10^{44}\,erg\,s^{-1}$ to examine whether this trend is also observed in less massive and more massive systems.

\begin{figure}
\centering
\begin{subfigure}{.45\textwidth}
  \centering
  \includegraphics[width=0.85\linewidth, height=5cm]{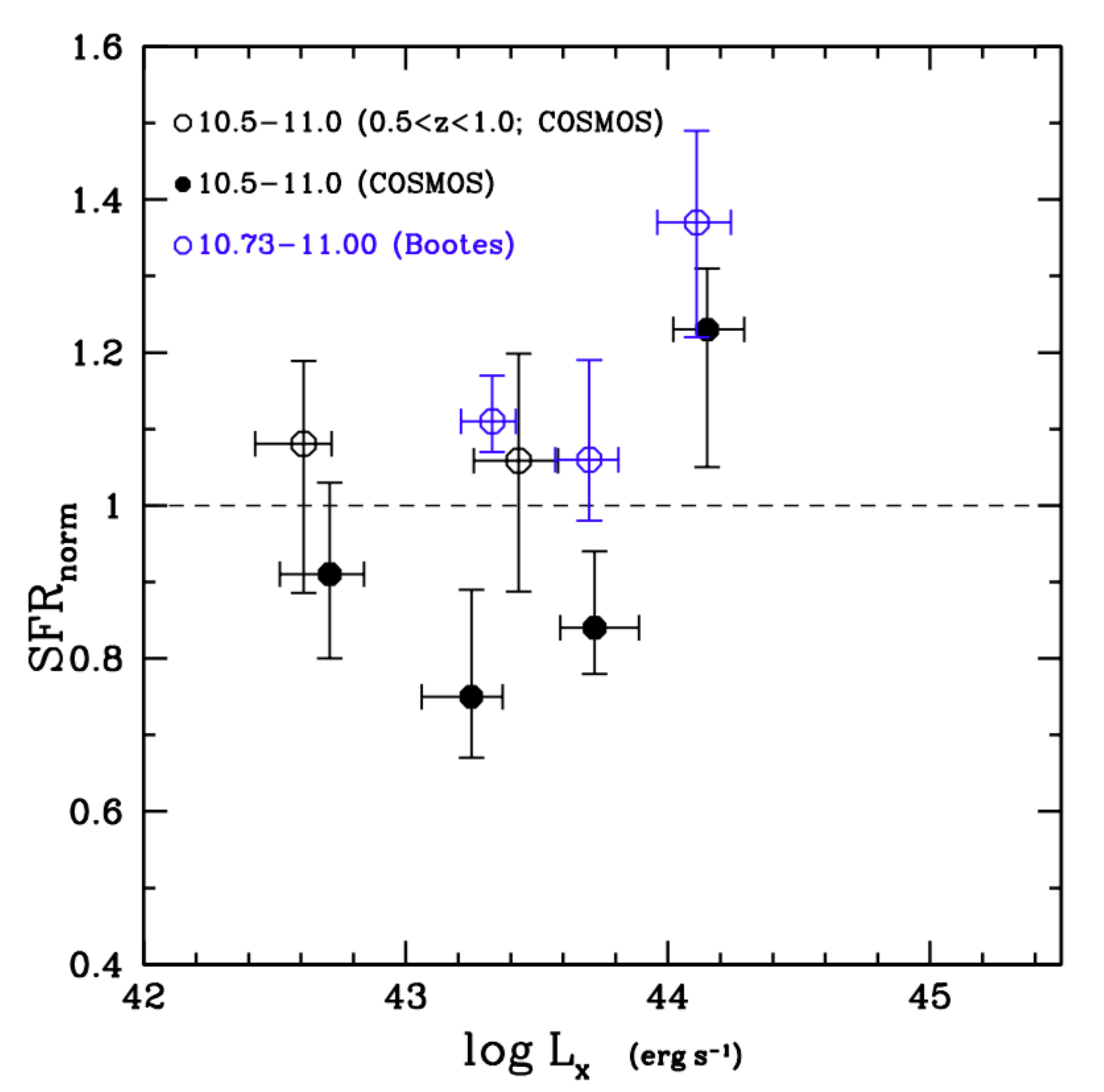}
  \label{}
\end{subfigure}
\begin{subfigure}{.45\textwidth}
  \centering
  \includegraphics[width=0.85\linewidth, height=5cm]{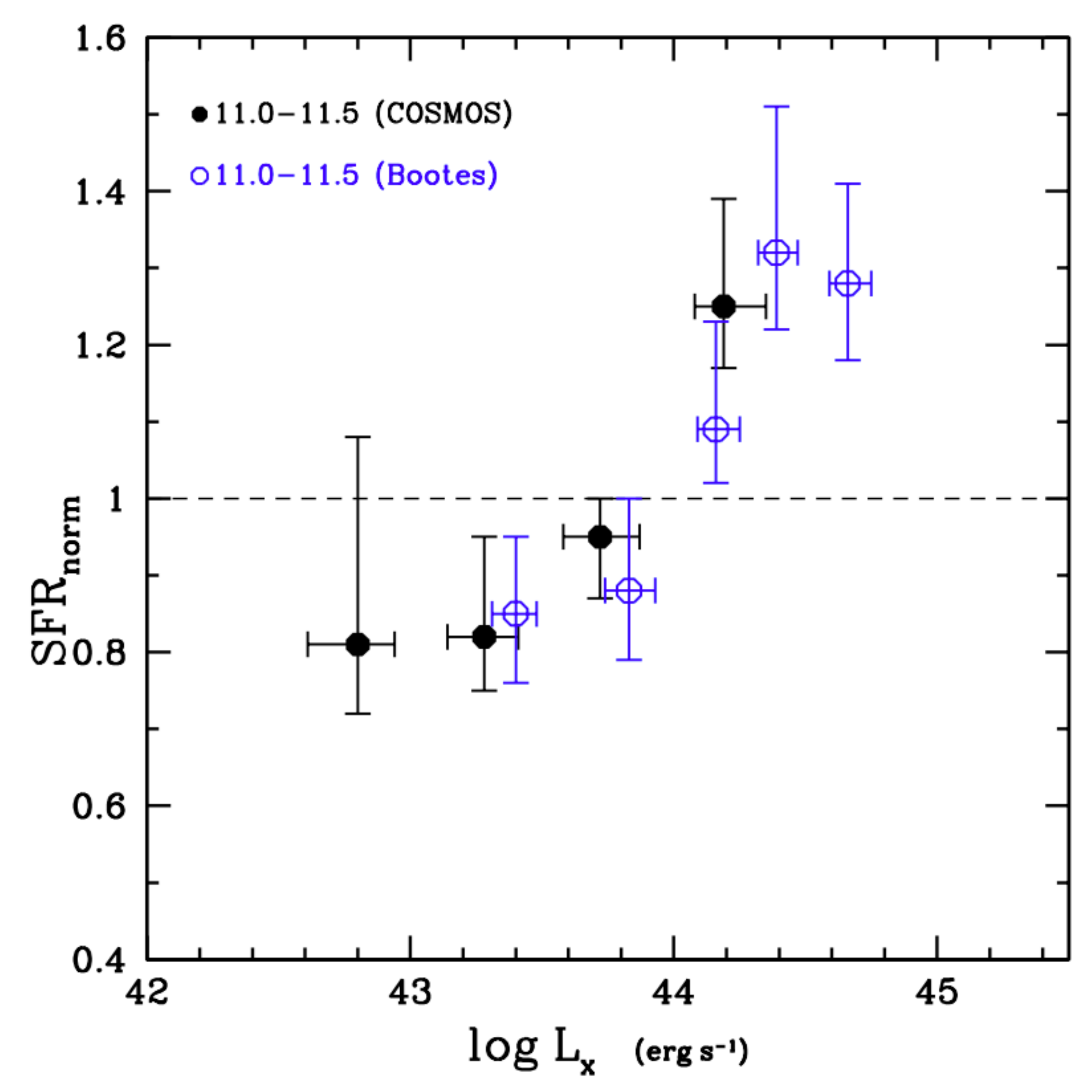}
  \label{}
\end{subfigure}
\begin{subfigure}{.45\textwidth}
  \centering
  \includegraphics[width=0.85\linewidth, height=5cm]{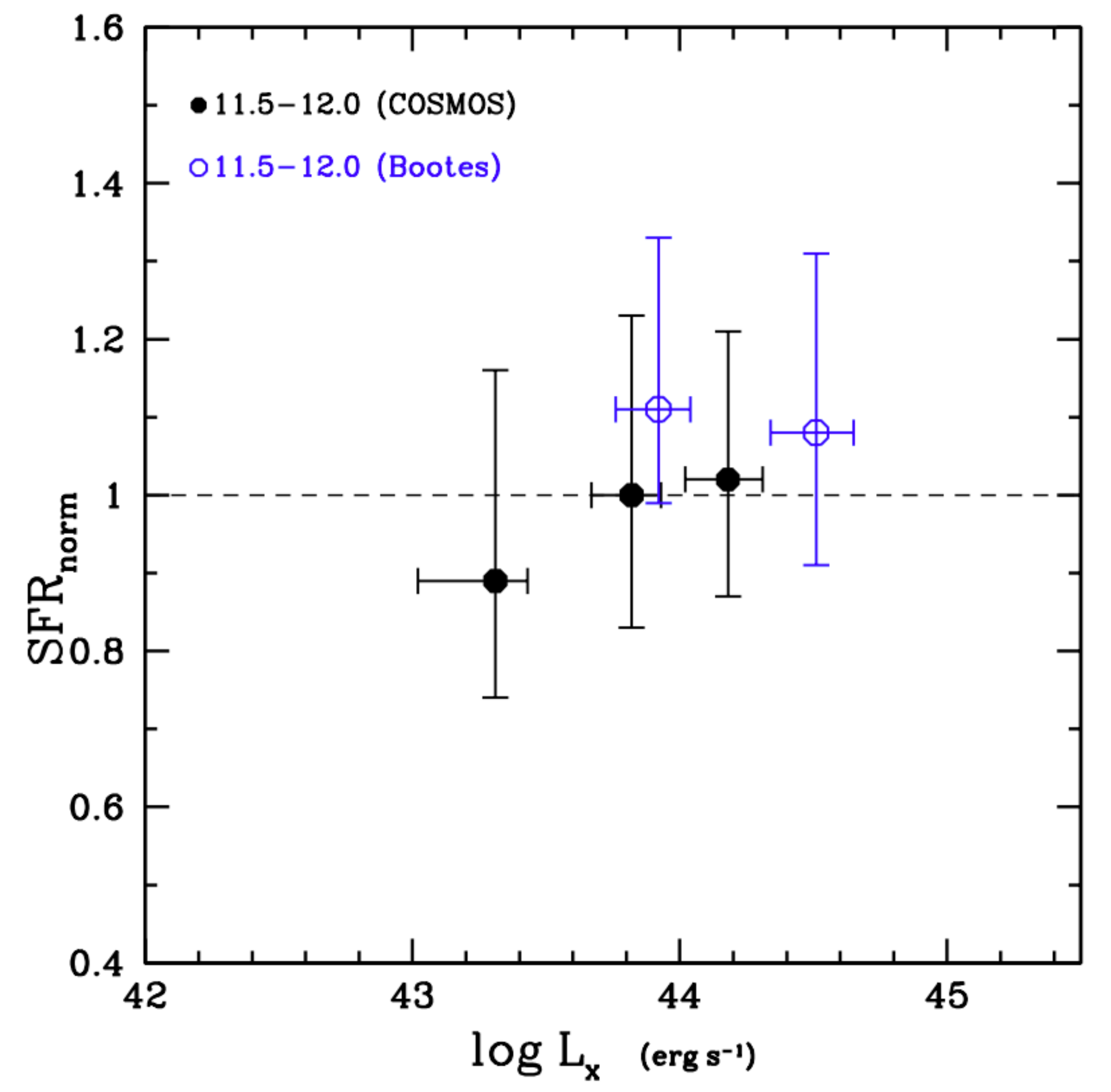}
  \label{}
\end{subfigure}
\caption{In this Figure, we complement the SFR$_{norm}$ values, in bins of L$_X$ and stellar mass, in the COSMOS field, with those found in the Bo$\rm \ddot{o}$tes field \citep{Mountrichas2021c}. We only present stellar mass intervals that are similar in the two studies. The small differences in the stellar mass bins are due to the different stellar mass completeness limits of the two surveys. Top panel: in the $\rm 10.5 < log\,[M_*(M_\odot)] < 11.0$ regime, both X-ray AGN in COSMOS and in Bo$\rm \ddot{o}$tes, present enhanced SFR$_{norm}$ compared to that at lower luminosities. The results in the Bo$\rm \ddot{o}$tes field (blue circles), include only sources at $\rm 0.5<z<1.0$ and appear higher compared to those in COSMOS (black circles). Open circles present the results in COSMOS for AGN with $\rm 0.5<z<1.0$ (open circles). The measurements are in statistical agreement with those at all redshifts (filled circles), but appear higher and in better agreement with those in Bo$\rm \ddot{o}$tes, in overlapping L$_X$. Middle panel: results for AGN that live in galaxies with $\rm 11.0 < log\,[M_*(M_\odot)] < 11.5$. The results in the two fields are in agreement. Most importantly, both present a similar increase of SFR$_{norm}$ at $\rm L_{X,2-10keV} > 10^{44}\,erg\,s^{-1}$. Bottom panel: measurements for AGN in the $\rm 11.5 < log\,[M_*(M_\odot)] < 12.0$ regime, for both fields. In this case, we do not detect enhanced SFR$_{norm}$ at high luminosities.}
\label{fig_sfrnorm_lx_mstar_bootes}
\end{figure}

In Fig. \ref{fig_sfrnorm_lx_mstar_bootes}, we complement our measurements in the COSMOS field with those in Bo$\rm \ddot{o}$tes from \cite{Mountrichas2021c}, in overlapping stellar mass intervals. Mountrichas et al. have presented their results in different redshift ranges (see their Fig. 11). For the comparison, we have re-calculated their SFR$_{norm}$ values, at all redshifts, taking into account the mass completeness at each redshift. The M$_*$ intervals in the case of the Bo$\rm \ddot{o}$tes dataset are slightly different, due to the mass completeness limits of the Bo$\rm \ddot{o}$tes dataset. In the $\rm 10.5 < log\,[M_*(M_\odot)] < 11.0$ regime, both X-ray AGN in COSMOS and in Bo$\rm \ddot{o}$tes, present enhanced SFR$_{norm}$ at $\rm L_{X,2-10keV} > 10^{44}\,erg\,s^{-1}$ compared to that at lower luminosities. The results in the Bo$\rm \ddot{o}$tes field (blue circles) though appear higher compared to those in COSMOS (black circles). We note that due to the mass completeness limits of Bo$\rm \ddot{o}$tes, only sources at $\rm 0.5<z<1.0$ contribute to this stellar mass interval. However, in the case of AGN in the COSMOS field, 30\% (82 out of 273) of AGN lie at $\rm 0.5<z<1.0$, in this stellar mass range. Although, our results in the previous section showed that SFR$_{norm}$ is independent of redshift, we also plot the results in the COSMOS field for AGN with $\rm 0.5<z<1.0$ (open circles). The measurements are in statistical agreement with those at all redshifts (filled circles), but appear higher and in better agreement with those in Bo$\rm \ddot{o}$tes, in overlapping L$_X$. Investigating this further, we find that the SFR distribution of sources in the reference catalogue at $\rm 0.5<z<1$ and with $\rm 10.5 < log\,[M_*(M_\odot)] < 11.0$, present a large tail at low SFR values, which could explain the increased SFR$_{norm}$ values found at this redshift interval. In the middle panel, we present the results for the two fields for AGN that live in galaxies with $\rm 11.0 < log\,[M_*(M_\odot)] < 11.5$. The results in both fields are in agreement. Most importantly, both present a similar increase of SFR$_{norm}$ at $\rm L_{X,2-10keV} > 10^{44}\,erg\,s^{-1}$. In the bottom panel we present, measurements for AGN in the $\rm 11.5 < log\,[M_*(M_\odot)] < 12.0$ regime, for both fields. In this case, we do not detect enhanced SFR$_{norm}$ at high luminosities. 

Overall, our results in COSMOS corroborate the findings in the Bo$\rm \ddot{o}$tes field, for increased SFR$_{norm}$ at  $\rm L_{X,2-10keV} > 10^{44}\,erg\,s^{-1}$ compared to lower L$_X$, for AGN that live in galaxies with $\rm 10.5 < log\,[M_*(M_\odot)] < 11.5$. A similar trend is not observed for more massive systems. This could be due to different physical mechanisms that fuel the SMBHs of the most massive systems compared to their lower mass counterparts. For example, in massive galaxies, diffuse gas could be accreted onto the SMBH without first being cooled onto the galaxies disk \citep[e.g.,][]{Fanidakis2013}, whereas in less massive galaxies AGN activity may be trigerred by galaxies mergers \citep[e.g.][]{Bower2006, Hopkins2008a} that also set off the star formation. Alternatively, the increase of SFR$_{norm}$ may occur at higher L$_X$ than those probed by our and the Bo$\rm \ddot{o}$tes datasets, for systems with $\rm log\,[M_*(M_\odot)] > 11.5$.

The trends we observe when we split our data into stellar mass bins (in addition to the luminosity bins) are in agreement with those found when we divide our sources only in L$_X$ bins (Sect. \ref{sfrnorm_lx_z}), but in the former case the trends are more evident. Specifically, at high luminosities AGN that reside in galaxies with $\rm 10.5 < log\,[M_*(M_\odot)] < 11.5$ seem to have enhanced SFR compared to MS galaxies of similar M$_*$. On the other hand, at lower L$_X$ our results show that the SFR of X-ray AGN is lower or, at most, equal to the SFR of MS galaxies.  

To further examine the correlation between SFR$_{norm}$, L$_X$ and M$_*$, we perform partial-correlation analysis (PCOR). PCOR measures the correlation between two variables while controlling for the effects of a third \citep[e.g.,][]{Lanzuisi2017, Yang2017, Fornasini2018}. We utilize one parametric statistic (Pearson) and one non-parametric statistic (Spearman). The results of the $p-$values are listed in Table \ref{table_pvalue}. The parametric method gives smaller $p-$values compared to the non-parametric method. This is because of the different assumptions made by the two methods \citep[for more details see Sect. 3.3 of][]{Yang2017}. Most importantly, $p-$values for the SFR$_{norm}$-M$_*$ relation are significantly smaller than the corresponding p-values for the SFR$_{norm}$-L$_X$ relation, suggesting that SFR$_{norm}$ correlates stronger with M$_*$ than with L$_X$, in agreement with the results presented above.

\begin{table}
\caption{$p-$ values of partial correlation analysis.}
\centering
\setlength{\tabcolsep}{2.mm}
\begin{tabular}{ccc}
& Pearson & Spearman  \\
  \hline
SFR$_{norm}$-L$_X$ & 0.519 & 0.726 \\  
SFR$_{norm}$-M$_*$ & 3.3$\times 10^{-7}$ & 2.3$\times 10^{-3}$\\

  \hline
\label{table_pvalue}
\end{tabular}
\end{table}

\begin{figure}
\centering
  \includegraphics[width=1.\linewidth, height=8.cm]{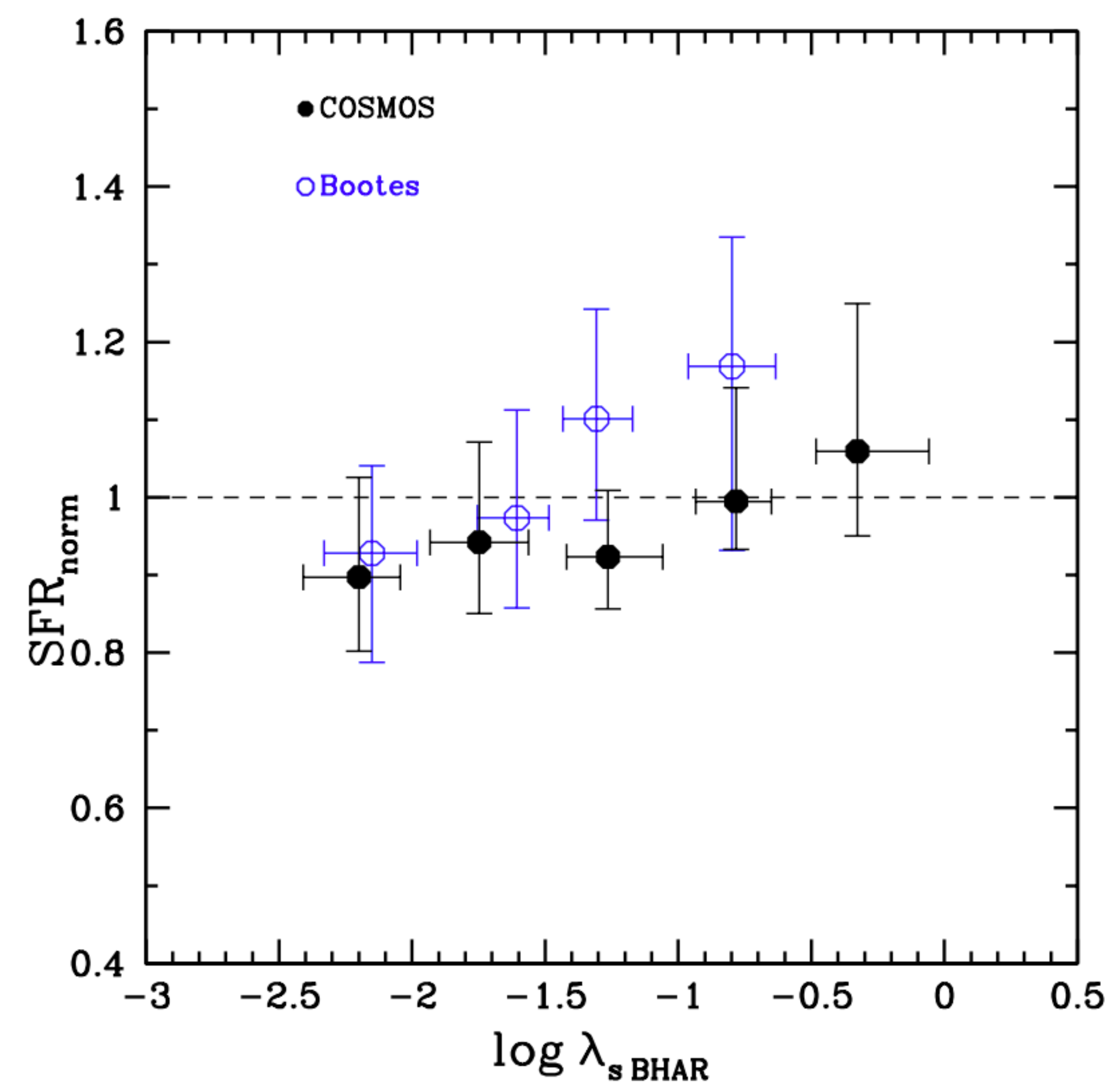}
  \caption{SFR$_{norm}$ vs. specific black hole accretion rate ($\lambda _{s, BHAR}$), for sources in COSMOS and in the Bo$\rm \ddot{o}$tes field. Median values are presented and the errors are calculated using bootstrap resampling. In both cases, we observe an increase of SFR$_{norm}$ with $\lambda _{s, BHAR}$. This is more evident in the case of the Bo$\rm \ddot{o}$tes dataset.} 
  \label{fig_lambda}
\end{figure}

\begin{table*}
\caption{Median values of $\rm log\,\lambda _{s\,BHAR}$, M$_*$, L$_X$ and redshift of X-ray AGN included in each of the $\rm log\,\lambda _{s\,BHAR}$ bin, presented in Fig. \ref{fig_lambda}, for sources in the COSMOS and Bo$\rm \ddot{o}$tes field \citep{Mountrichas2021c}.}
\centering
\setlength{\tabcolsep}{2.mm}
\begin{tabular}{cccccc}
& \multicolumn{5}{c}{COSMOS / Bo$\rm \ddot{o}$tes}   \\
  \hline
$\rm log\,\lambda _{s\,BHAR}$ & $-2.18$ / $-2.11$ & $-1.75$ / $-1.65$ & $-1.26$ / $-1.28$ & $-0.81$ / $-0.83$ & $-0.33$ / --- \\
  \hline
$\rm log\,[M_*(M_\odot)]$ & 11.28 / 11.45 &11.19 / 11.36& 10.99 / 11.33 & 10.72 / 11.27 & 10.38 / --- \\
$\rm log\,(L_{X,2-10keV}/(ergs^{-1}))$ &43.04 / 43.27  & 43.44 / 43.65 & 43.75 / 44.07 & 44.00 / 44.64  & 44.00 / --- \\
redshift & 0.91 / 0.79 & 1.20 / 0.97& 1.59 / 1.17 & 1.75 / 1.33  &1.73 / --- \\
  \hline
\label{table_lambda}
\end{tabular}
\end{table*}

\subsection{SFR$_{norm}$ as a function of specific black hole accretion rate}
\label{sec_sfrnorm_lambda}

{\cite{Mountrichas2021c} examined the relation between SFR$_{norm}$ and the $\lambda_{s, BHAR}$ parameter, which quantifies the rate of accretion onto the SMBH relative to the M$_*$ of the host galaxy. Their results showed a linear increase of SFR$_{norm}$ with $\lambda_{s, BHAR}$. We follow their analysis and compare our calculations using the COSMOS dataset with their measurements. For the estimation of $\lambda_{s, BHAR}$, we use the analytical expression (2) of \cite{Aird2018}:

\begin{equation}
\lambda_{s,BHAR}=\frac{k_{bol} \times L_X}{1.3\times 10^{38}\rm {erg\,s^{-1}}\times 0.002\frac{M_*}{M\odot}},
\end{equation}
where $k_{bol}$ is a bolometric correction factor. We adopt the same value as in \cite{Aird2018}, i.e. $k_{bol}=25$. Fig. \ref{fig_lambda}, presents the results of our measurements and those from \cite{Mountrichas2021c}, in the Bo$\rm \ddot{o}$tes field. For the latter, we have not use the averaged values presented in their Fig. 10 (black circles). Instead, we have calculated their SFR$_{norm}$ values, at all redshifts spanned by their dataset, taking into account the mass completeness at each redshift. Our results show a mild increase of SFR$_{norm}$ with $\lambda_{s, BHAR}$}. However, this increase is not as evident as that seen using the Bo$\rm \ddot{o}$tes sample. This is confirmed by the expressions that describe the best fit of the measurements. In the case of COSMOS sources, SFR$_{norm}=\rm 0.080^{+0.015}_{-0.052}log\,\lambda _{s, BHAR}+1.064^{+0.148}_{-0.204}$ compared to SFR$_{norm}=\rm 0.189^{+0.029}_{-0.037}log\,\lambda _{s, BHAR}+1.320^{+0.162}_{-0.181}$ for Bo$\rm \ddot{o}$tes. Moreover, with the exception of the first two $\lambda_{s, BHAR}$ bins, SFR$_{norm}$ values of COSMOS AGN appear consistent, but lower than those of the Bo$\rm \ddot{o}$tes sources.

In Table \ref{table_lambda}, we present the median $\rm log\,\lambda _{s\,BHAR}$, M$_*$, L$_X$ and redshift of X-ray AGN included in each of the $\rm log\,\lambda _{s\,BHAR}$ bins, presented in Fig. \ref{fig_lambda}. Comparing the properties of the AGN between the two datasets, we find that COSMOS sources are less luminous and less massive than their Bo$\rm \ddot{o}$tes counterparts, in all $\rm log\,\lambda _{s\,BHAR}$ bins. This become more evident as we move to higher $\rm log\,\lambda _{s\,BHAR}$ values. This could explain the results presented in Fig. \ref{fig_lambda}. The AGN included in the first two $\rm log\,\lambda _{s\,BHAR}$ bins, have similar properties (L$_X$, M$_*$) and similar SFR$_{norm}$ values. For the next two $\rm log\,\lambda _{s\,BHAR}$bins, the SFR$_{norm}$ is higher in the case of AGN in Bo$\rm \ddot{o}$tes compared to those in COSMOS, but in this case Bo$\rm \ddot{o}$tes X-ray sources are more luminous and significantly more massive.


\subsection{Comparison with previous studies}

In this section, we compare our results with recent studies. \cite{Masoura2021} found a strong dependence of the SFR$_{norm}$ with L$_X$ (left panel of their Fig. 10) and evolution of the SFR$_{norm}$-L$_X$ relation with redshift (left panel of their Fig. 11). Although, we do not find the strong dependence of the SFR$_{norm}$ with L$_X$ found in their work, both studies agree that AGN tend to have lower SFR compared to MS galaxies at  $\rm L_{X,2-10keV} < 10^{44}\,erg\,s^{-1}$, while at $\rm L_{X,2-10keV} \sim 10^{44}\,erg\,s^{-1}$ AGN have similar SFR with MS galaxies. \cite{Masoura2021} results suggest that at higher L$_X$ AGN have enhanced SFR compared to star forming MS galaxies. Although, our results show indications for enhanced SFR of AGN compared to the SFR of the galaxy reference catalogue, in particular for AGN hosts with $\rm 10.5 < log\,[M_*(M_\odot)] < 11.5$, the luminosity baseline covered by our COSMOS sample complemented by the Bo$\rm \ddot{o}$tes sources used in \cite{Mountrichas2021c} does not extend to sufficiently high L$_X$ and with sufficiently large number of X-ray sources, to allow a fair comparison. We leave the comparison at these high luminosities for a future paper that we make use of the recently released eROSITA X-ray catalogue. 

Moreover, our results do not show evolution of the SFR$_{norm}$-L$_X$ with redshift, in contrast to the findings of Masoura et al. However, an important difference in the analysis used in the two works, is that in \cite{Masoura2021}, they compare the SFR of AGN with that of MS galaxies, using for the latter the analytical expression of \cite{Schreiber2015}. As it was shown in \cite{Mountrichas2021c} this approach introduces different systematics at different redshift intervals (see their Fig. 6). Therefore, at least to some degree,  the different amplitude of the SFR$_{norm}$-L$_X$ found at different redshifts in \cite{Masoura2021} could be attributed to these systematic effects. 

\cite{Bernhard2019} used 541 X-ray AGN in the COSMOS field, within $\rm 0.8<z<1.2$. They estimate SFR$_{norm}$, using the \cite{Schreiber2015} analytical formula. They found that higher luminosity AGN $\rm L_{X,2-10keV} > 2 \times 10^{43}\,erg\,s^{-1}$ present a narrower SFR$_{norm}$ distribution compared to their lower L$_X$ counterparts. The former is also shifted to higher values, closer to that of MS galaxies. Although the two studies cannot be directly compared due to the different analysis followed (cf calculation of SFR of MS galaxies), they are in broad agreement in the sense that higher L$_X$ AGN seem to have SFR closer to the MS than their lower L$_X$ counterparts.

A more fair comparison can be made with \cite{Santini2012}. In that study they used X-ray selected AGN in the GOODS-S, GOODS-N and XMM-COSMOS and compared the SFR of X-ray sources with a mass-match control sample of non-AGN galaxies. Their analysis showed that  the star formation of AGN in the COSMOS field (bottom panel of their Fig. 5) is consistent with the star formation of MS galaxies at redshifts $\rm 0.5<z<2.5$, in agreement with our findings.

\section{Summary-Conclusions}
\label{sec_summary}

We used $\sim 1000$ X-ray sources in the $\it{COSMOS}-Legacy$ survey \citep{Marchesi2016} within the UltraVISTA region and compared their SFR  with $\sim 90,000$ non-AGN systems compiled by the HELP collaboration \citep{Shirley2019, Shirley2021}, over a wide range of redshifts ($\rm 0<z<2.5$) and X-ray luminosities ($\rm 10^{42.5}< L_{X,2-10keV} < 10^{44.0}\,erg\,s^{-1}$). The galaxy control sample has been observed in the same field as the X-ray sources and the same photometric selection criteria have been applied on both datasets. We performed SED fitting, using X-CIGALE, with the same parametric grid on both datasets to measure (host) galaxy properties. The same mass completeness limits were applied on both samples. These enabled us to avoid systematics and compare the SFR of the two populations in a uniform and consistent manner.

First, we studied the SFR-L$_X$ relation with respect to the position of the host galaxy to the MS, at different luminosity and redshift intervals. Our results showed no evolution of the SFR-L$_X$ with redshift. The SFR of AGN was found lower or, at most, equal to the SFR of MS galaxies at $\rm L_{X,2-10keV} < 10^{44}\,erg\,s^{-1}$. We complemented our results with those in the Bo$\rm \ddot{o}$tes field \citep{Mountrichas2021c}, since the same analysis has been performed in both works. This allows us to extend the luminosity baseline to higher values. We observed a mild increase of the SFR of AGN at $\rm L_{X,2-10keV} > 2-3 \times 10^{44}\,erg\,s^{-1}$. At this luminosity regime, AGN appear to have enhanced SFR compared to non-AGN systems.  

Prompted by previous studies \citep[e.g.,][]{Mountrichas2021c}, we examined the SFR as a function of L$_X$ for different stellar mass ranges. The results are consistent with those found when we divided the sources only in L$_X$ bins. However, when the stellar mass is taken into consideration the trends are more evident. Our results also corroborate the findings in the  Bo$\rm \ddot{o}$tes field. Specifically, the enhanced AGN SFR is detected for AGN that live in galaxies with $\rm 10.5 < log\,[M_*(M_\odot)] < 11.5$. A similar trend is not observed for more massive systems. 

Overall, our results combined with those in the Bo$\rm \ddot{o}$tes field show a very mild dependence of SFR$_{norm}$  with L$_X$, which becomes (more) evident when we take into account the stellar mass of the AGN host galaxies. Specifically, at low and moderate luminosities ($\rm 10^{42}< L_{X,2-10keV} < 10^{44}\,erg\,s^{-1}$) AGN tend to have lower or, at most, similar SFR with star forming MS galaxies. At   $\rm L_{X,2-10keV} \sim 2-3\times 10^{44}\,erg\,s^{-1}$  and $\rm 10.5 < log\,[M_*(M_\odot)] < 11.5$, AGN present enhanced SFR compared to non-AGN systems.

Although the data used in this work, complemented by those in the Bo$\rm \ddot{o}$tes field, span two orders of magnitude in X-ray luminosity, they do not cover the highest L$_X$ regime ($\rm L_{X,2-10keV} > 10^{44.5}\,erg\,s^{-1}$). AGN at higher L$_X$ will allow us to populate the SFR-L$_X$ plots with data points at high luminosities and examine whether the SFR of AGN keeps increasing as we move to even higher luminosities ($\rm L_{X,2-10keV} > 10^{45}\,erg\,s^{-1}$). Ongoing and future surveys (eROSITA, {\it{Athena}}) will provide us with a large number of the most powerful X-ray sources and allow us to answer these questions.

\begin{acknowledgements}
GM acknowledges support by the Agencia Estatal de Investigación, Unidad de Excelencia María de Maeztu, ref. MDM-2017-0765.
VAM acknowledges support by the Grant RTI2018-096686-B-C21 funded by MCIN/AEI/10.13039/501100011033 and by 'ERDF A way of making Europe'.
This research is co-financed by Greece and the European Union (European Social Fund-ESF) through the Operational Programme ''Human Resources Development, Education and Lifelong Learning 2014-2020'' in the context of the project ''Anatomy of galaxies: their stellar and dust content through cosmic time'' (MIS 5052455). The project has received funding from Excellence Initiative of Aix-Marseille University - AMIDEX, a French 'Investissements d'Avenir' programme.

\end{acknowledgements}

\bibliography{mybib}{}

\begin{thebibliography}{64}
\expandafter\ifx\csname natexlab\endcsname\relax\def\natexlab#1{#1}\fi

\bibitem[{Aird {et~al.}(2018)Aird, Coil, \& Georgakakis}]{Aird2018}
Aird, J., Coil, A.~L., \& Georgakakis, A. 2018, Monthly Notices of the Royal
  Astronomical Society, 474, 1225

\bibitem[{Aird {et~al.}(2019)Aird, Coil, \& Georgakakis}]{Aird2019}
Aird, J., Coil, A.~L., \& Georgakakis, A. 2019, Monthly Notices of the Royal
  Astronomical Society, 484, 4360

\bibitem[{{Arnouts} {et~al.}(1999){Arnouts}, {Cristiani}, {Moscardini},
  {Matarrese}, {Lucchin}, {Fontana}, \& {Giallongo}}]{Arnouts1999}
{Arnouts}, S., {Cristiani}, S., {Moscardini}, L., {et~al.} 1999, MNRAS, 310,
  540

\bibitem[{Bernhard {et~al.}(2019)Bernhard, Grimmett, Mullaney, Daddi,
  Tadhunter, \& Jin}]{Bernhard2019}
Bernhard, E., Grimmett, L.~P., Mullaney, J.~R., {et~al.} 2019, Monthly Notices
  of the Royal Astronomical Society: Letters, 483, L52

\bibitem[{Boquien {et~al.}(2019)Boquien, Burgarella, Roehlly, Buat, Ciesla,
  Corre, Inoue, \& Salas}]{Boquien2019}
Boquien, M., Burgarella, D., Roehlly, Y., {et~al.} 2019, Astronomy {\&}
  Astrophysics, 622, A103

\bibitem[{Bower {et~al.}(2012)Bower, Benson, \& Crain}]{Bower2012}
Bower, R.~G., Benson, A.~J., \& Crain, R.~A. 2012, MNRAS, 422, 2816

\bibitem[{{Bower} {et~al.}(2006){Bower}, {Benson}, {Malbon}, {Helly}, {Frenk},
  {Baugh}, {Cole}, \& {Lacey}}]{Bower2006}
{Bower}, R.~G., {Benson}, A.~J., {Malbon}, R., {et~al.} 2006, MNRAS, 370, 645

\bibitem[{Bruzual \& Charlot(2003)}]{Bruzual_charlot2003}
Bruzual, G. \& Charlot, S. 2003, MNRAS, 344, 1000

\bibitem[{Buat {et~al.}(2019)Buat, Ciesla, Boquien, Ma{\l}ek, \&
  Burgarella}]{Buat2019}
Buat, V., Ciesla, L., Boquien, M., Ma{\l}ek, K., \& Burgarella, D. 2019,
  Astronomy {\&} Astrophysics, 632, A79

\bibitem[{Buat {et~al.}(2021)Buat, Mountrichas, Yang, Boquien, Roehlly,
  Burgarella, Stalevski, Ciesla, \& Theul{\'{e}}}]{Buat2021}
Buat, V., Mountrichas, G., Yang, G., {et~al.} 2021, A\&A, 654, A93

\bibitem[{{Cappelluti} {et~al.}(2009)}]{Cappelluti2009}
{Cappelluti}, N. {et~al.} 2009, A\&A, 497, 635

\bibitem[{Charlot \& Fall(2000)}]{Charlot_Fall_2000}
Charlot, S. \& Fall, S.~M. 2000, ApJ, 539, 718

\bibitem[{Civano {et~al.}(2016)Civano, Marchesi, Comastri, Urry, Elvis,
  Cappelluti, Puccetti, Brusa, Zamorani, Hasinger, Aldcroft, Alexander,
  Allevato, Brunner, Capak, Finoguenov, Fiore, Fruscione, Gilli, Glotfelty,
  Griffiths, Hao, Harrison, Jahnke, Kartaltepe, Karim, LaMassa, Lanzuisi,
  Miyaji, Ranalli, Salvato, Sargent, Scoville, Schawinski, Schinnerer,
  Silverman, Smolcic, Stern, Toft, Trakhenbrot, Treister, \&
  Vignali}]{Civano2016}
Civano, F., Marchesi, S., Comastri, A., {et~al.} 2016, ApJ, 819, 62

\bibitem[{{Dale} {et~al.}(2014){Dale}, {Helou}, {Magdis}, {Armus},
  {D{\'{\i}}az-Santos}, \& {Shi}}]{Dale2014}
{Dale}, D.~A., {Helou}, G., {Magdis}, G.~E., {et~al.} 2014, ApJ, 784, 83

\bibitem[{Dubois {et~al.}(2016)Dubois, Peirani, Pichon, Devriendt, Gavazzi,
  Welker, \& Volonteri}]{Dubois2016}
Dubois, Y., Peirani, S., Pichon, C., {et~al.} 2016, MNRAS, 463, 3948

\bibitem[{Dunn \& Fabian(2006)}]{Dunn2006}
Dunn, R. J.~H. \& Fabian, A.~C. 2006, MNRAS, 373, 959

\bibitem[{Elbaz {et~al.}(2007)Elbaz, Daddi, Borgne, Dickinson, Alexander,
  Chary, Starck, Brandt, Kitzbichler, MacDonald, Nonino, Popesso, Stern, \&
  Vanzella}]{Elbaz2007}
Elbaz, D., Daddi, E., Borgne, D.~L., {et~al.} 2007, Astronomy {\&}
  Astrophysics, 468, 33

\bibitem[{Fanidakis {et~al.}(2013)Fanidakis, Georgakakis, Mountrichas, Krumpe,
  Baugh, Lacey, Frenk, Miyaji, \& Benson}]{Fanidakis2013}
Fanidakis, N., Georgakakis, A., Mountrichas, G., {et~al.} 2013, Monthly Notices
  of the Royal Astronomical Society, 435, 679

\bibitem[{Florez {et~al.}(2020)Florez, Jogee, Sherman, Stevans, Finkelstein,
  Papovich, Kawinwanichakij, Ciardullo, Gronwall, Urry, Kirkpatrick, LaMassa,
  Ananna, \& Wold}]{Florez2020}
Florez, J., Jogee, S., Sherman, S., {et~al.} 2020, Monthly Notices of the Royal
  Astronomical Society, 497, 3273

\bibitem[{Fornasini {et~al.}(2018)Fornasini, Civano, Fabbiano, Elvis, Marchesi,
  Miyaji, \& Zezas}]{Fornasini2018}
Fornasini, F.~M., Civano, F., Fabbiano, G., {et~al.} 2018, The Astrophysical
  Journal, 865, 43

\bibitem[{{Georgakakis} {et~al.}(2017)}]{Georgakakis2017}
{Georgakakis}, A. {et~al.} 2017, MNRAS, 469, 3232

\bibitem[{{Hickox} \& {Alexander}(2018)}]{Hickox2018}
{Hickox}, R.~C. \& {Alexander}, D.~M. 2018, ARAA, 56, 625

\bibitem[{Hickox {et~al.}(2014)Hickox, Mullaney, Alexander, Chen, Civano, \&
  Goulding}]{Hickox2014}
Hickox, R.~C., Mullaney, J.~R., Alexander, D.~M., {et~al.} 2014, ApJ, 782, 11

\bibitem[{{Hopkins} {et~al.}(2008){Hopkins}, {Hernquist}, {Cox}, \&
  {Keres}}]{Hopkins2008a}
{Hopkins}, P.~F., {Hernquist}, L., {Cox}, T.~J., \& {Keres}, D. 2008, ApJS,
  175, 356

\bibitem[{{Ilbert} {et~al.}(2006)}]{Ilbert2006}
{Ilbert}, O. {et~al.} 2006, A\&A, 457, 841

\bibitem[{Koutoulidis {et~al.}(2021)Koutoulidis, Mountrichas, Georgantopoulos,
  Pouliasis, \& Plionis}]{Koutoulidis2021}
Koutoulidis, L., Mountrichas, G., Georgantopoulos, I., Pouliasis, E., \&
  Plionis, M. 2021, A\&A accepted [\eprint{2111.02539}]

\bibitem[{Laigle {et~al.}(2016)Laigle, McCracken, Ilbert, Hsieh, Davidzon,
  Capak, Hasinger, Silverman, Pichon, Coupon, Aussel, Borgne, Caputi, Cassata,
  Chang, Civano, Dunlop, Fynbo, Kartaltepe, Koekemoer, F{\`{e}}vre, Floc'h,
  Leauthaud, Lilly, Lin, Marchesi, Milvang-Jensen, Salvato, Sanders, Scoville,
  Smolcic, Stockmann, Taniguchi, Tasca, Toft, Vaccari, \& Zabl}]{Laigle2016}
Laigle, C., McCracken, H.~J., Ilbert, O., {et~al.} 2016, ApJS, 224, 24

\bibitem[{{Lanzuisi} {et~al.}(2017)}]{Lanzuisi2017}
{Lanzuisi}, G. {et~al.} 2017, A\&A, 602, 13

\bibitem[{Loh(2008)}]{Loh2008}
Loh, J.~M. 2008, ApJ, 681, 726

\bibitem[{{Lutz} {et~al.}(2010)}]{Lutz2010}
{Lutz}, D. {et~al.} 2010, ApJ, 712, 1287

\bibitem[{Marchesi {et~al.}(2016)Marchesi, Civano, Elvis, Salvato, Brusa,
  Comastri, Gilli, Hasinger, Lanzuisi, Miyaji, Treister, Urry, Vignali,
  Zamorani, Allevato, Cappelluti, Cardamone, Finoguenov, Griffiths, Karim,
  Laigle, LaMassa, Jahnke, Ranalli, Schawinski, Schinnerer, Silverman, Smolcic,
  Suh, \& Trakhtenbrot}]{Marchesi2016}
Marchesi, S., Civano, F., Elvis, M., {et~al.} 2016, ApJ, 817, 34

\bibitem[{Masoura {et~al.}(2021)Masoura, Mountrichas, Georgantopoulos, \&
  Plionis}]{Masoura2021}
Masoura, V.~A., Mountrichas, G., Georgantopoulos, I., \& Plionis, M. 2021,
  Astronomy {\&} Astrophysics, 646, A167

\bibitem[{Masoura {et~al.}(2018)Masoura, Mountrichas, Georgantopoulos, Ruiz,
  Magdis, \& Plionis}]{Masoura2018}
Masoura, V.~A., Mountrichas, G., Georgantopoulos, I., {et~al.} 2018, A\&A, 618,
  31

\bibitem[{McCracken {et~al.}(2012)McCracken, Milvang-Jensen, Dunlop, Franx,
  Fynbo, F{\`{e}}vre, Holt, Caputi, Goranova, Buitrago, Emerson, Freudling,
  Hudelot, L{\'{o}}pez-Sanjuan, Magnard, Mellier, M{\o}ller, Nilsson,
  Sutherland, Tasca, \& Zabl}]{McCracken2012}
McCracken, H.~J., Milvang-Jensen, B., Dunlop, J., {et~al.} 2012, A\&A, 544,
  A156

\bibitem[{Morganti(2017)}]{Morganti2017}
Morganti, R. 2017, Nature Astronomy, 1, 39

\bibitem[{Mountrichas {et~al.}(2021{\natexlab{a}})Mountrichas, Buat,
  Georgantopoulos, Yang, Masoura, Boquien, \& Burgarella}]{Mountrichas2021b}
Mountrichas, G., Buat, V., Georgantopoulos, I., {et~al.} 2021{\natexlab{a}},
  Astronomy {\&} Astrophysics, 653, A70

\bibitem[{Mountrichas {et~al.}(2021{\natexlab{b}})Mountrichas, Buat, Yang,
  Boquien, Burgarella, \& Ciesla}]{Mountrichas2021a}
Mountrichas, G., Buat, V., Yang, G., {et~al.} 2021{\natexlab{b}}, Astronomy
  {\&} Astrophysics, 646, A29

\bibitem[{Mountrichas {et~al.}(2021{\natexlab{c}})Mountrichas, Buat, Yang,
  Boquien, Burgarella, Ciesla, Malek, \& Shirley}]{Mountrichas2021c}
Mountrichas, G., Buat, V., Yang, G., {et~al.} 2021{\natexlab{c}}, Astronomy
  {\&} Astrophysics, 653, A74

\bibitem[{Mullaney {et~al.}(2015)Mullaney, Alexander, Aird, Bernhard, Daddi,
  Moro, Dickinson, Elbaz, Harrison, Juneau, Liu, Pannella, Rosario, Santini,
  Sargent, Schreiber, Simpson, \& Stanley}]{Mullaney2015}
Mullaney, J.~R., Alexander, D.~M., Aird, J., {et~al.} 2015, Monthly Notices of
  the Royal Astronomical Society: Letters, 453, L83

\bibitem[{Noeske {et~al.}(2007)Noeske, Weiner, Faber, Papovich, Koo,
  Somerville, Bundy, Conselice, Newman, Schiminovich, Floch, Coil, Rieke, Lotz,
  Primack, Barmby, Cooper, Davis, Ellis, Fazio, Guhathakurta, Huang, Kassin,
  Martin, Phillips, Rich, Small, Willmer, \& Wilson}]{Noeske2007}
Noeske, K.~G., Weiner, B.~J., Faber, S.~M., {et~al.} 2007, The Astrophysical
  Journal, 660, L43

\bibitem[{{Page} {et~al.}(2012)}]{Page2012}
{Page}, M.~J. {et~al.} 2012, nat, 485, 213

\bibitem[{Park {et~al.}(2006)Park, Kashyap, Siemiginowska, van Dyk, Zezas,
  Heinke, \& Wargelin}]{Park2006}
Park, T., Kashyap, V.~L., Siemiginowska, A., {et~al.} 2006, The Astrophysical
  Journal, 652, 610

\bibitem[{{Pozzetti} {et~al.}(2010)}]{Pozzetti2010}
{Pozzetti}, L. {et~al.} 2010, A\&A, 523, 23

\bibitem[{Rosario {et~al.}(2013)Rosario, Trakhtenbrot, Lutz, Netzer, Trump,
  Silverman, Schramm, Lusso, Berta, Bongiorno, Brusa, Förster-Schreiber,
  Genzel, Lilly, Magnelli, Mainieri, Maiolino, Merloni, Mignoli, Nordon,
  Popesso, Salvato, Santini, Tacconi, \& Zamorani}]{Rosario2013}
Rosario, D.~J., Trakhtenbrot, B., Lutz, D., {et~al.} 2013, Astronomy {\&}
  Astrophysics, 560, A72

\bibitem[{{Salvato} {et~al.}(2011)}]{Salvato2011}
{Salvato}, M. {et~al.} 2011, ApJ, 742, 61

\bibitem[{Santini {et~al.}(2012)Santini, Rosario, Shao, Lutz, Maiolino,
  Alexander, Altieri, Andreani, Aussel, Bauer, Berta, Bongiovanni, Brandt,
  Brusa, Cepa, Cimatti, Daddi, Elbaz, Fontana, Schreiber, Genzel, Grazian,
  Floc'h, Magnelli, Mainieri, Nordon, Garcia, Poglitsch, Popesso, Pozzi,
  Riguccini, Rodighiero, Salvato, Sanchez-Portal, Sturm, Tacconi, Valtchanov,
  \& Wuyts}]{Santini2012}
Santini, P., Rosario, D.~J., Shao, L., {et~al.} 2012, Astronomy {\&}
  Astrophysics, 540, A109

\bibitem[{Schreiber {et~al.}(2015)}]{Schreiber2015}
Schreiber, C. {et~al.} 2015, A\&A, 575, 29

\bibitem[{{Scoville} {et~al.}(2007)}]{Scoville2007}
{Scoville}, N. {et~al.} 2007, ApJS, 172, 1

\bibitem[{Shimizu {et~al.}(2015)Shimizu, Mushotzky, Mel{\'{e}}ndez, Koss, \&
  Rosario}]{Shimizu2015}
Shimizu, T.~T., Mushotzky, R.~F., Mel{\'{e}}ndez, M., Koss, M., \& Rosario,
  D.~J. 2015, Monthly Notices of the Royal Astronomical Society, 452, 1841

\bibitem[{Shimizu {et~al.}(2017)Shimizu, Mushotzky, Mel{\'{e}}ndez, Koss,
  Barger, \& Cowie}]{Shimizu2017}
Shimizu, T.~T., Mushotzky, R.~F., Mel{\'{e}}ndez, M., {et~al.} 2017, Monthly
  Notices of the Royal Astronomical Society, 466, 3161

\bibitem[{Shirley {et~al.}(2021)Shirley, Duncan, Varillas, Hurley, Ma{\l}ek,
  Roehlly, Smith, Aussel, Bakx, Buat, Burgarella, Christopher, Duivenvoorden,
  Eales, Efstathiou, Solares, Griffin, Jarvis, Faro, Marchetti, McCheyne,
  Papadopoulos, Penner, Pons, Prescott, Rigby, Rottgering, Saxena, Scudder,
  Vaccari, Wang, \& Oliver}]{Shirley2021}
Shirley, R., Duncan, K., Varillas, M. C.~C., {et~al.} 2021, MNRAS, 507, 129

\bibitem[{Shirley {et~al.}(2019)Shirley, Roehlly, Hurley, Buat, del Carmen
  Campos~Varillas, Duivenvoorden, Duncan, Efstathiou, Farrah, Solares, Malek,
  Marchetti, McCheyne, Papadopoulos, Pons, Scipioni, Vaccari, \&
  Oliver}]{Shirley2019}
Shirley, R., Roehlly, Y., Hurley, P.~D., {et~al.} 2019, Monthly Notices of the
  Royal Astronomical Society, 490, 634

\bibitem[{Speagle {et~al.}(2014)Speagle, Steinhardt, Capak, \&
  Silverman}]{Speagle2014}
Speagle, J.~S., Steinhardt, C.~L., Capak, P.~L., \& Silverman, J.~D. 2014, The
  Astrophysical Journal Supplement Series, 214, 15

\bibitem[{Stalevski {et~al.}(2012)Stalevski, Fritz, Baes, Nakos, \&
  Popovi{\'{c}}}]{Stalevski2012}
Stalevski, M., Fritz, J., Baes, M., Nakos, T., \& Popovi{\'{c}}, L.~{\v{C}}.
  2012, Monthly Notices of the Royal Astronomical Society, 420, 2756

\bibitem[{Stalevski {et~al.}(2016)Stalevski, Ricci, Ueda, Lira, Fritz, \&
  Baes}]{Stalevski2016}
Stalevski, M., Ricci, C., Ueda, Y., {et~al.} 2016, Monthly Notices of the Royal
  Astronomical Society, 458, 2288

\bibitem[{{Sutherland} \& {Saunders}(1992)}]{Sutherland_and_Saunders1992}
{Sutherland}, W. \& {Saunders}, W. 1992, MNRAS, 259, 413

\bibitem[{Torbaniuk {et~al.}(2021)Torbaniuk, Paolillo, Carrera, Cavuoti,
  Vignali, Longo, \& Aird}]{Torbaniuk2021}
Torbaniuk, O., Paolillo, M., Carrera, F., {et~al.} 2021, MNRAS, 506, 2619

\bibitem[{Volonteri {et~al.}(2015)Volonteri, Capelo, Netzer, Bellovary, Dotti,
  \& Governato}]{Volonteri2015}
Volonteri, M., Capelo, P.~R., Netzer, H., {et~al.} 2015, Monthly Notices of the
  Royal Astronomical Society, 449, 1470

\bibitem[{Weaver {et~al.}(2021)Weaver, Kauffmann, Ilbert, McCracken, Moneti,
  Toft, Brammer, Shuntov, Davidzon, Hsieh, Laigle, Anastasiou, Jespersen,
  Vinther, Capak, Casey, McPartland, Milvang-Jensen, Mobasher, Sanders,
  Zalesky, Arnouts, Aussel, Dunlop, Faisst, Franx, Furtak, Fynbo, Gould, Greve,
  Gwyn, Kartaltepe, Kashino, Koekemoer, Kokorev, Fevre, Lilly, Masters, Magdis,
  Mehta, Peng, Riechers, Salvato, Sawicki, Scarlata, Scoville, Shirley,
  Sneppen, Smolcic, Steinhardt, Stern, Tanaka, Taniguchi, Teplitz, Vaccari,
  Wang, \& Zamorani}]{Weaver2021}
Weaver, J.~R., Kauffmann, O.~B., Ilbert, O., {et~al.} 2021, AJSS accepted
  [\eprint{2110.13923}]

\bibitem[{Whitaker {et~al.}(2012)Whitaker, van Dokkum, Brammer, \&
  Franx}]{Whitaker2012}
Whitaker, K.~E., van Dokkum, P.~G., Brammer, G., \& Franx, M. 2012, ApJ, 754,
  L29

\bibitem[{Yang {et~al.}(2022)Yang, Boquien, Brandt, Buat, Burgarella, Ciesla,
  Lehmer, Małek, Mountrichas, Papovich, Pons, Stalevski, Theulé, \&
  Zhu}]{Yang2022}
Yang, G., Boquien, M., Brandt, W.~N., {et~al.} 2022, A\&A [\eprint{2201.03718}]

\bibitem[{Yang {et~al.}(2020)Yang, Boquien, Buat, Burgarella, Ciesla, Duras,
  Stalevski, Brandt, \& Papovich}]{Yang2020}
Yang, G., Boquien, M., Buat, V., {et~al.} 2020, Monthly Notices of the Royal
  Astronomical Society, 491, 740

\bibitem[{Yang {et~al.}(2018)Yang, Brandt, Vito, Chen, Trump, Luo, Sun, Xue,
  Koekemoer, Schneider, Vignali, \& Wang}]{Yang2018}
Yang, G., Brandt, W.~N., Vito, F., {et~al.} 2018, Monthly Notices of the Royal
  Astronomical Society, 475, 1887

\bibitem[{{Yang} {et~al.}(2017){Yang}, {Chen}, {Vito}, {Brandt}, {Alexander},
  {Luo}, {Sun}, {Xue}, {Bauer}, {Koekemoer}, {Lehmer}, {Liu}, {Schneider},
  {Shemmer}, {Trump}, {Vignali}, \& {Wang}}]{Yang2017}
{Yang}, G., {Chen}, C. T.~J., {Vito}, F., {et~al.} 2017, ApJ, 842, 72

\end{thebibliography}
\bibliographystyle{aa}

\end{document}